\documentclass[envcountsame]{llncs}
\usepackage{xcolor}
\usepackage{hyperref}
\usepackage{mathtools}
\usepackage{multirow}
\usepackage{stmaryrd}
\usepackage{cancel}
\usepackage{amsmath,amssymb}
\usepackage{thmtools, thm-restate}
\usepackage[shortlabels]{enumitem}
\usepackage{microtype}
\usepackage{pbox}
\usepackage{marvosym}
\usepackage[noline,noend,linesnumbered]{algorithm2e}
\usepackage{cite}
\hypersetup{
	pdftitle={Improving Automatic Complexity Analysis of Integer Programs}, colorlinks=true, linkcolor=blue, citecolor=olive, filecolor=magenta, urlcolor=cyan
}
\usepackage{todonotes}
\usepackage{placeins}
\usepackage{tikz}
\usepackage{caption}
\usepackage{subcaption}
\usetikzlibrary{shapes,calc,arrows,automata,decorations.pathmorphing}
\usepackage[capitalize,nameinlink]{cleveref}

\newenvironment{myproof}{\noindent{\it Proof.}}{\qed\medskip}
\usepackage[location=appendix,prependtoappendix=true]{moveproofs}
\renewcommand{\epsilon}{\varepsilon}
 \let\oldvarphi\varphi \renewcommand{\phi}{\oldvarphi}
\renewcommand{\emptyset}{\varnothing}

\newcommand{\oldcomment}[1]{}
\newcommand{\delete}[1]{}

\crefname{definition}{Def.}{Def.}
\crefname{example}{Ex.}{Ex.}
\crefname{appendix}{App.}{App.}
\crefname{ex}{Ex.}{Ex.}
\crefname{theorem}{Thm.}{Thm.}
\crefname{lemma}{Lemma}{Lemma}
\crefname{section}{Sect.}{Sect.}
\crefname{subsection}{Sect.}{Sect.}
\crefname{algorithm}{Alg.}{Alg.}
\crefname{corollary}{Cor.}{Cor.}
\crefname{figure}{Fig.}{Fig.}

\input{insbox}

\usepackage{wrapfig}

\makeatletter \newcommand*\bigcdot{\mathpalette\bigcdot@{.4}}
\newcommand*\bigcdot@[2]{\mathbin{\vcenter{\hbox{\scalebox{#2}{$\m@th#1\bullet$}}}}}
\makeatother

\newcommand{\NN}{\mathbb{N}}
\newcommand{\ZZ}{\mathbb{Z}}
\newcommand{\QQ}{\mathbb{Q}}
\newcommand{\R}{\mathbb{R}}
\newcommand{\C}{\mathbb{C}}
\newcommand{\MRF}{\text{M}\Phi\text{RF}}
\newcommand{\MRFs}{\text{M}\Phi\text{RFs}}
\newcommand{\linpoly}[1]{\ZZ[#1]_{\mathrm{lin}}}

\newcommand{\var}{\texttt}

\newcommand{\true}{\var{true}}

\newcommand{\location}{\ell}
\newcommand{\valuation}{\sigma}
\newcommand{\Valuation}{\Sigma}
\newcommand{\update}{\eta}

\newcommand{\guard}{\tau}

\newcommand{\landau}{\mathcal{O}}

\newcommand{\abs}[1]{\left|#1\right|}

\newcommand{\Size}{{\mathcal{SB}}}

\newcommand{\Time}{{\mathcal{RB}}}

\newcommand{\prestate}{{\tilde{\valuation}}}
\newcommand{\actstate}{\valuation}

\newcommand{\prel}{{\tilde{\location}}}
\newcommand{\actl}{\location}

\newcommand{\RV}{\mathcal{RV}}
\newcommand{\BoundSet}{\mathcal{B}}
\newcommand{\Program}{\mathcal{P}}

\newcommand{\ConstraintSet}{\mathcal{C}}
\newcommand{\TSet}{\mathcal{T}}
\newcommand{\Set}{\mathcal{S}}

\newcommand{\VSet}{\mathcal{V}}

\newcommand{\TVSet}{\mathcal{TV}}
\newcommand{\PVSet}{\mathcal{PV}}

\newcommand{\LSet}{\mathcal{L}}

\newcommand{\maximum}[1]{\max \left\lbrace #1 \right\rbrace}

\newcommand{\braced}[1]{\lbrace #1 \rbrace}

\newcommand{\overapprox}[1]{\left\lceil #1 \right\rceil}

\newcommand{\exacteval}[2]{#2\left(#1\right)}
\newcommand{\upeval}[2]{\exacteval{#1}{\abs{#2}}}

\newcommand{\timeboundterm}[1][t]{
	\sup \braced{k \in \NN \mid \location \in \LSet, \valuation\in \Valuation, (\location_0, \valuation_0) \, (\rightarrow^* \circ \rightarrow_{#1})^k \, (\location, \valuation)}
}

\newcommand{\KoAT}[0]{\tool{KoAT}}
\DeclareMathOperator{\rc}{rc}
\newcommand{\mdepth}{\mathit{mdepth}}

\newcommand{\tool}[1]{\textsf{#1}}

\newcommand{\CITS}[0]{\textsf{Complexity\_ITS}}
\newcommand{\CCINT}[0]{\textsf{Complexity\_C\_Integer}}
\newcommand{\CITSsh}[0]{\textsf{CITS}}
\newcommand{\CCINTsh}[0]{\textsf{CINT}}

\newcommand{\automaticcomplanalysis}[0]{\cite{dp-framework,aprove,koat,dblp:conf/rta/avanzinim13,dblp:conf/tacas/avanzinims16,dblp:journals/toplas/0002ah12,loopus,cofloco2,cofloco3,costa-complexity,maxcore,campy,rank,pubs,c4b,pubs-upper-lower,ramlpopl17,aprove-java,dblp:journals/iandc/mosers18,dblp:journals/sosym/albertbghpr16}}

\newcommand{\noopsort}[1]{}

\title{Improving Automatic Complexity Analysis of Integer Programs\thanks{funded by Deutsche Forschungsgemeinschaft (DFG, German Research Foundation) - 235950644 (Project GI 274/6-2) and  DFG Research Training Group 2236 UnRAVeL}
}
\author{J\"urgen Giesl
  \and Nils Lommen
  \and Marcel Hark
	\and Fabian Meyer
}
\institute{
	LuFG Informatik 2, RWTH Aachen University, Germany
	\email{\{giesl,lommen,marcel.hark,fabian.meyer\}@cs.rwth-aachen.de}
}

\begin{document}
\allowdisplaybreaks
\setlength{\abovedisplayskip}{3pt}
\setlength{\belowdisplayskip}{3pt}
\setlength{\abovedisplayshortskip}{1pt}
\setlength{\belowdisplayshortskip}{3pt}
\maketitle \begin{abstract}
	In \cite{koat}, we developed an approach for automatic complexity analysis of integer programs, based on an alternating modular inference of upper runtime and size bounds for program parts.
	In this paper, we show how recent techniques to improve automated termination analysis of integer programs (like the generation of multiphase-linear ranking functions and control-flow refinement) can be integrated into our approach for the inference of runtime bounds.
	The power of the resulting approach is demonstrated by an extensive experimental evaluation with our new re-implementation of the corresponding tool \KoAT.
\end{abstract}

\FloatBarrier
\section{Introduction}
\label{sec:intro}
\InsertBoxR{4}{
	\begin{minipage}{0.32\linewidth}
		\centering \LinesNotNumbered \DontPrintSemicolon
		\begin{algorithm}[H]
			\While{$x > 0$}{
				$x \gets x + y$\;
				$y \gets y - 1$\;
			}
		\end{algorithm}
		\vspace*{-.1cm}
		\captionof{figure}{Loop without Linear Ranking Function}\label{fig:linear}
		\vspace*{1.8cm}
	\end{minipage}
}
\noindent
There are many techniques and tools for automated complexity analysis of pro\-grams, e.g., \automaticcomplanalysis.
Most of them infer variants of (mostly linear) poly\-nomial ranking functions (see, e.g.,\linebreak \cite{podelski2004complete,bradley_polyranking}) which are then combined to get a runtime bound for the overall program.
However, approaches based on linear ranking functions are incomplete for termination and thus also for complexity analysis.
For example, consider the loop from \cite{genaim_mrf,leike_linear} in \cref{fig:linear},
 which terminates, but does not admit a linear ranking function.
Its runtime is linear in the initial values of $x$ and $y$, if they are positive initially.
The reason is that if $y > 0$, then $x$ grows first but it is decreased with the same ``speed'' once $y$ has become negative.

Recently so-called multiphase-linear ranking functions have gained interest (see, e.g., \cite{genaim_mrf,leike_linear,genaim_mrf_recurrent,dblp:journals/sttt/yuanls21}).
For loops as in \cref{fig:linear}, ranking functions of this form detect that the program has two phases:\ first $y$ is decremented until it is negative.
Afterwards, $x$ is decremented until it is negative and the loop terminates.
In \cite{genaim_mrf}, it is shown that the existence of a multiphase-linear ranking function for a loop implies linear runtime complexity.
In the current paper, we embed multiphase-linear ranking functions into our modular approach for complexity analysis of integer programs from \cite{koat}.
In contrast to \cite{genaim_mrf}, we infer multiphase-linear ranking functions for parts of the program and combine the so-obtained bounds to an overall runtime bound.
In this way, \pagebreak  we obtain a powerful technique which is able to infer finite runtime bounds for programs that contain loops such as \cref{fig:linear}.

Moreover, different forms of control-flow refinement were used to improve the automatic termination and complexity analysis of programs further, see, e.g., \cite{cofloco3,domenech2019control}.
The basic idea is to gain ``more information'' on the values of variables to sort out certain paths in the program.
For example, the control-flow refinement technique from \cite{domenech2019control} detects that the programs in \cref{fig:crf} and \cref{fig:unrolled} are equivalent.
Clearly, the program in \cref{fig:unrolled} is easier to analyze as the two consecutive loops do not interfere with each other: $x$ and $z$ are constants in its first loop, while $y$ and $z$ are constants in its second loop.
We show how to integrate the technique for control-flow refinement from \cite{domenech2019control} into our modular analysis in a non-trivial way.
This increases the power of our approach further.

\begin{figure}[t]
	\centering \hspace*{-.7cm}
	\begin{minipage}[t]{0.3\linewidth}
		\LinesNotNumbered
		\begin{algorithm}[H]
			\While{$x < 0$}{
				\eIf {$y < z$}{
					$y \gets y - x$
				}{
					$x \gets x + 1$
				}
			}
		\end{algorithm}
		\vspace*{-.3cm}
		\captionof{figure}{Original Loop}\label{fig:crf}
	\end{minipage}
	\hspace*{.8cm}
	\begin{minipage}[t]{0.5\linewidth}
		\LinesNotNumbered
		\begin{algorithm}[H]
			\hspace*{.5cm}	\While{$x < 0 \wedge y < z$}{
				\hspace*{.5cm}	$y \gets y - x$
			}
			\hspace*{.5cm}	\While{$x < 0 \wedge y \geq z$}{
				\hspace*{.5cm}	$x \gets x + 1$
			}
		\end{algorithm}
		\hspace*{1cm}\\
		\vspace*{-.3cm}
		\captionof{figure}{After Control-Flow Refinement}\label{fig:unrolled}
	\end{minipage}
	\vspace*{-.5cm}
\end{figure}

\paragraph{Structure:}
We first recapitulate our approach from \cite{koat} in \cref{sec:preliminaries}.
Afterwards, we adapt it to multiphase-linear ranking functions in \cref{sec:mrf}.
In \cref{sec:cfr}, we discuss how to incorporate control-flow refinement from \cite{domenech2019control} into our analysis.
We provide an extensive experimental evaluation of our corresponding new version of the tool \tool{KoAT} \cite{koatwebsite} and compare it with existing tools in \cref{sec:eval}.
Finally, we discuss related work and conclude (\cref{sec:conclusion}).
All proofs can be found in \cref{app:proofs}.

\FloatBarrier
\section{Preliminaries}
\label{sec:preliminaries}
In this section we recapitulate our approach for complexity analysis from
\cite{koat}.
We first introduce \emph{constraints}, which are used in the guards of programs.

\begin{definition}[Constraints]
	Let $\VSet$ be a set of variables.
	The set of \emph{constraints} $\ConstraintSet(\VSet)$ over $\VSet$ is the smallest set containing $e_1 \leq e_2$ for all polynomials $e_1,e_2 \in \ZZ[\VSet]$ and $c_1 \land c_2$ for all $c_1,c_2\in\ConstraintSet(\VSet)$.
\end{definition}
In addition to ``$\leq$'', we also use relations like ``$>$'' and ``$=$'', which can be simulated by constraints (e.g., $e_1 > e_2$ is equivalent to $e_2 +1 \leq e_1$ when regarding integers).

Now we define the notion of integer programs which we use in this paper.
Instead of \textbf{while} loops as in \Cref{fig:linear,fig:crf,fig:unrolled}, we use a formalism based on transitions (which of course also allows us to represent \textbf{while} programs easily).

\begin{definition}[Integer Program]
	\label{def:integer_program}
	An \emph{integer program} $\Program$ over a set of variables $\VSet$ is a tuple $(\PVSet,\LSet,\location_0,\TSet)$ of
	\begin{itemize}
		\item[$\bullet$] a finite set of \emph{program variables} $\PVSet \subseteq \VSet$,
		\item[$\bullet$] a finite set of \emph{locations} $\LSet$ with a distinguished \emph{initial location} $\location_0\in\LSet$, and
		\item[$\bullet$] a finite set of \emph{transitions} $\TSet$.
			A transition is a tuple $(\location,\guard,\update,\location')$
	consisting  of
			\begin{enumerate}
				\item the \emph{start location} $\location \in \LSet$ and
	the  \emph{target location} $\location' \in \LSet\setminus\{\location_0\}$,
				\item the \emph{guard} $\guard\in\ConstraintSet(\VSet)$ of
	$t$, \pagebreak and
				\item the \emph{update function}
				      $\eta\colon\PVSet\rightarrow \ZZ[\VSet]$ of $t$, mapping every program variable to an update polynomial.
			\end{enumerate}
	\end{itemize}
	We call $\TVSet = \VSet \setminus \PVSet$ the set of \emph{temporary variables}.
\end{definition}
Note that the initial location has \emph{no} incoming transitions.
The transitions $(\location_0, ...)$ whose start location is $\location_0$ are called \emph{initial} transitions.

Thus, integer programs contain two kinds of non-determinism.
Non-determi\-nistic branching is realized by multiple transitions with the same start location whose guards are non-exclusive.
Non-deterministic sampling is modeled by temporary variables (which can be restricted in the guard of a transition).
Temporary variables are not updated in the program.
Intuitively, these variables are set by an adversary trying to ``sabotage'' the program in order to obtain long runtimes.

\begin{example}
	Consider the integer program in \cref{fig:running} over the program variables $\PVSet = \{x,y,z\}$, the locations $\LSet = \{ \location_0, \location_1, \location_2 \}$, and the transitions $\TSet = \{
		t_0, t_1, t_2, t_3 \}$. In \Cref{fig:running}, we omitted trivial guards, i.e., $\tau = \true$, and trivial updates, i.e., updates of the form $\eta(v)=v$.
	This integer program corresponds to two nested loops: the inner loop is given by $t_2$, the outer loop by $t_1$ and $t_3$.

	\InsertBoxR{0}{
		\begin{minipage}{0.7\linewidth}
			\vspace*{.1cm}
			\begin{tikzpicture}[->,>=stealth',shorten >=1pt,auto,node distance=2.5cm,semithick,initial text=$ $]
				\node[state,initial] (q0) {$\location_0$}; \node[state] (q1) [right of=q0,xshift=-1cm]{$\location_1$}; \node[state] (q2) [right of=q1]{$\location_2$}; \draw (q0) edge node [text width=.5cm,align=center] {$t_0$} (q1) (q1) edge[bend left] node [text width=3cm,align=center] {$t_1: \guard = z > 0$ \\
						$\update(x) = z-1 $ \\
						$\update(y) = z-1$} (q2) (q2) edge[loop right] node [text width=3cm,align=center] {\hspace*{-.8cm}$t_2:\guard = x > 0$ \\
						\hspace*{-.8cm}$\update(x) = x+y $ \\
						\hspace*{-.8cm}$\update(y) = y-1$} (q2) (q2) edge[bend left] node [text width=3cm,align=center] {$t_3:\guard = z > 0$ \\
						$\update(z) = z-1 $
					} (q1);
			\end{tikzpicture}
			\vspace*{-.7cm}
			
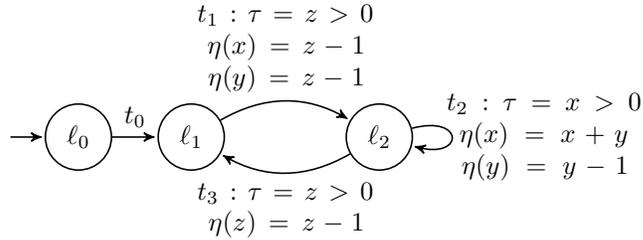
\captionof{figure}{Integer Program with Nested Loops}
			\label{fig:running}
			\vspace*{2.8cm}
		\end{minipage}
	}	Transition $t_0$ just forwards the input values.
	If $z>0$, $t_1$ sets $x$ and $y$ to $z-1$.
	Then, $t_2$ decrements $y$ by $1$ and updates $x$ to $x+y$ repeatedly as long as $x > 0$ (i.e., it corresponds to the loop in \cref{fig:linear}).
	Transition $t_3$ decrements $z$ by $1$ and leads back to the starting point of the outer loop.

	Note that $t_2$ and $t_3$ correspond to a non-deterministic branching as their guards are non-exclusive.
	If $t_0$ had the update $\eta(x) = u$ and the guard $u>0$, then this would correspond to a non-deterministic sampling of a positive value.
\end{example}

From now on, we fix an integer program $\Program$ over the variables $\VSet$.
A mapping $\valuation: \VSet \rightarrow \ZZ$ is called a \emph{state} and $\Valuation$ denotes the set of all states.
We also apply states to arithmetic expressions $e$ and constraints $c$, where the number $\valuation(e)$ resp.\ the Boolean value $\valuation(c)$ results from $e$ resp.\ $c$ by replacing each variable $v$ by $\valuation(v)$.

\begin{definition}[Evaluation of Integer Programs]
	A \emph{configuration} is an element of $\LSet \times \Valuation$.
	For two configurations $(\location,\valuation)$ and $(\location',\valuation')$, and a transition $t = (\location_t,\guard,\update,\location_{t}')\in\TSet$, $(\location,\valuation)\rightarrow_t(\location',\valuation')$ is an \emph{evaluation}
	step by $t$ if
	\begin{itemize}
		\item[$\bullet$] $\location = \location_t$ and $\location' = \location_{t}'$,
		\item[$\bullet$] $\valuation(\guard) = \normalfont{\true}$, and
		\item[$\bullet$] for every program variable $v\in\PVSet$ we have $\valuation(\update(v)) = \valuation'(v)$.
	\end{itemize}
	We denote the union of all relations $\to_t$ for
	$t \in \TSet$
        by $\to_{\TSet}$.
	Whenever it is clear  from the context, we omit the transition $t$ resp.\ the set $\TSet$ in the index.
	We also \pagebreak abbreviate $(\location_0,\valuation_0)\rightarrow_{t_1}(\location_1,\valuation_1) \cdots \rightarrow_{t_k}(\location_k,\valuation_k)$ by $(\location_0,\valuation_0)\rightarrow^k(\location_k,\valuation_k)$.
\end{definition}

\begin{example}
	For the integer program in \cref{fig:running}, when denoting program states $\valuation$ as tuples $(\valuation(x),\valuation(y),\valuation(z)) \in \ZZ^3$,
	we have $(\location_0,(0,0,2)) \to_{t_0}(\location_1,(0,0,2)) \to_{t_1}
		(\location_2,(1,1,2))\to_{t_2}(\location_2,(2,0,2))\to_{t_3}(\location_1,(2,0,1))$.
\end{example}

For an integer program, the (worst-case) runtime complexity w.r.t.\ an initial state $\valuation_0$ is defined to be the length of the longest evaluation starting in $\valuation_0$.

\begin{definition}[Runtime Complexity]
	The (worst-case) \emph{runtime complexity} of $\Program$ is the function $\rc: \Valuation \rightarrow \overline{\NN}$ with $\overline{\NN}
		= \NN \cup\{\omega\}$ and $\rc(\valuation_0) = \sup\{k \in \NN \mid\linebreak
		\location_k \in \LSet,\valuation_k \in \Valuation, (\location_0,\valuation_0) \to^{k} (\location_k,\valuation_k)\}$ for all $\valuation_0 \in \Valuation$.
\end{definition}

As in \cite{koat}, our approach combines bounds for program parts.
We restrict ourselves to bounds that represent weakly monotonically increasing functions.
Such bounds have the advantage that they can easily be ``composed'', i.e., if $f$ and $g$ are both weakly monotonically increasing upper bounds, then so is $f \circ g$.

\begin{definition}[Bounds]
	\label{def:bound}
	The set of \emph{bounds}
	$\BoundSet$ is the smallest set with $\overline{\NN} \subseteq \BoundSet$, $\PVSet \subseteq \BoundSet$, $b_1+b_2 \in \BoundSet$, $b_1 \cdot b_2 \in \BoundSet$, and $k^b \in \BoundSet \text{ for all } k \in \NN$ and $b,b_1,b_2 \in \BoundSet$.

	A bound which is only constructed from $\NN$, $\PVSet$, $+$, and $\cdot$ is called \emph{polynomial}.
	A polynomial bound of degree at most $1$ is called \emph{linear}.
\end{definition}

For any $\valuation \in \Valuation$, $\abs{\valuation}$ denotes the state with $\abs{\valuation}(v) = \abs{\valuation(v)}$ for all $v \in \VSet$.
Clearly, a bound $b \in \BoundSet$ induces a weakly monotonic function on states by mapping any $\valuation \in \Valuation$ to $\abs{\valuation}(b) \in \overline{\NN}$.
Then, $\abs{\valuation} \leq \abs{\valuation'}$ implies $\abs{\valuation}(b) \leq \abs{\valuation'}(b)$.
As usual, we compare\linebreak
functions pointwise, i.e., $\abs{\valuation} \leq \abs{\valuation'}$ means that $\abs{\valuation}(v) \leq \abs{\valuation'}(v)$ for all $v \in \VSet$.

\begin{example}
	\label{ex:bounds}
	For $\PVSet = \{x,y\}$, we have $\omega$, $x^2$, $x+y$, $2^{x^2+y} \in \BoundSet$.
	Here, $x^2$ and $x+y$ are polynomial bounds and $x+y$ is linear.
	Consider the state $\valuation$ with $\valuation(x) = 1$ and $\valuation(y) = -2$.
	Then, $\abs{\valuation}\left(x+y\right) = \abs{1} + \abs{-2} = 3$.
\end{example}

To over-approximate the runtime complexity, we now introduce the concepts of\linebreak runtime and size bounds.
A runtime bound for a transition $t\!\in\!\TSet$ over-approximates\linebreak the maximal number of occurrences of that transition in any evaluation starting with the initial state $\valuation_0 \in \Valuation$.
Here, $\rightarrow^* \circ \rightarrow_t$ denotes the relation describing arbitrary many evaluation steps followed by a step with transition $t$.

\begin{definition}[Runtime Bound]
	\label{def:runtime_bound}
	The function $\Time: \TSet \rightarrow \BoundSet$ is a \emph{runtime bound} if for all $t \in \TSet$ and all states $\valuation_0 \in \Valuation$ we have
	\[
		\abs{\valuation_0}\left(\Time(t)\right) \; \geq \; \sup \braced{k \in \NN \mid \location \in \LSet, \valuation\in \Valuation, (\location_0, \valuation_0) \, (\rightarrow^* \circ \rightarrow_t)^k \, (\location, \valuation)}.
	\]
\end{definition}
Note that we require the runtime bound to only contain \emph{program variables} since the values of temporary variables are ``set by the adversary''.

\begin{example}
	\label{ex:time}
	For the program in \cref{fig:running}, the technique from \cite{koat}
	obtains the following runtime bound.
	Trivially, $\Time(t_0)=1$, as $t_0$ can only be applied once in any evaluation.
	Since the outer loop is only executed if $z>0$ and every iteration of the outer loop decreases $z$ by $1$, we get $\Time(t_1)=\Time(t_3) = z$, i.e., these transitions can occur at most $\abs{z_0}$ times, if $z$ has the value $z_0 \in \ZZ$ initially.
	However, the implementation of \cite{koat} in the original version of the 
	tool \KoAT{} cannot infer a finite runtime bound for $t_2$ since this transition
	does not admit a linear ranking function, i.e.,  \pagebreak a linear function which
	decreases by at least one and is bounded from below for each iteration of the loop.
	Intuitively, the reason is that $x$ is bounded, but it does not decrease in every iteration.
	In contrast, $y$ decreases in every iteration, but it is not bounded.
	In \Cref{sec:mrf}, we will show how to improve our approach for complexity analysis such that it obtains a finite runtime bound for transitions like $t_2$ (see \cref{ex:mrf_final}).
\end{example}

The following corollary shows that every runtime bound $\Time$ directly yields an upper bound for the program's runtime complexity: Instead of over-approximating the runtime complexity of the full program at once, one can compute runtime bounds for each transition separately and simply add these bounds.

\begin{corollary}[Over-Approximating $\rc$]
	\label{thm:over_approx_rc}
	Let $\Time$ be a runtime bound.
	Then for all states $\valuation_0 \in \Valuation$ we have $\upeval{\sum\nolimits_{t \in \TSet} \Time(t)}{\valuation_0} \; \geq \; \rc(\valuation_0)$.
\end{corollary}

The framework in \cite{koat} performs a \emph{modular} analysis of the program, i.e., parts of the program are analyzed as standalone programs and the results are then lifted to contribute to the overall analysis.
For example, for the integer program in \cref{fig:running}, the inner loop $t_2$ is analyzed separately in order to compute its runtime bound.
But to lift a local runtime bound of $t_2$ to a runtime bound of $t_2$ in the full program, one has to take into account that the values of the variables when executing $t_2$ are not the input values of the program, but the values that the variables have after an execution of the previous transition $t_1$.

So to compute the runtime bound of a transition $t'$, our approach considers all transitions $t$ that can occur directly before $t'$ in evaluations and it needs size bounds $\Size(t,v)$ to over-approximate the absolute values that the variables $v \in \PVSet$ may have \emph{after} these ``previous'' transitions $t$.
(This intuition will later be formalized in \cref{thm:time_bound}.)
Here, we call $\RV = \TSet \times \PVSet$ the set of \emph{result variables}.

\begin{definition}[Size Bound]
	\label{def:size_bound}
	The function $\Size: \RV \rightarrow \BoundSet$ is a \emph{size bound} if for all
	$(t,v) \in \RV$ and all states $\valuation_0 \in \Valuation$ we have
	\[
		\abs{\valuation_0}\left(\Size(t,v)\right) \; \geq \; \sup \braced{\abs{\valuation(v)} \in \NN \mid \location \in \LSet, \valuation\in \Valuation, (\location_0, \valuation_0) \, (\rightarrow^* \circ \rightarrow_t) \, (\location, \valuation)}.
	\]
\end{definition}

\begin{example}
	\label{ex:size}
	Consider again the program in \cref{fig:running}.
	Here, $\Size(t_0,v) = v$ for $v \in \{x,y,z\}$, because $t_0$ does not change any variable.
	So if $(\location_0, \valuation_0) \rightarrow_{t_0} (\location_1, \valuation_1)$ then $\abs{\valuation_0}\left(\Size(t_0,v)\right) = \abs{\valuation_0}(v) = \abs{\valuation_1}(v)$.
	Moreover, $\Size(t_1,z) = \Size(t_2,z) = \Size(t_3,z) = z$ as $z$ is never increased in the program.
	For the computation of $\Size(t_1,x)$ and $\Size(t_1,y)$, the approach of \cite{koat} sums up the values of $\Size(t_0,z)$ and $\Size(t_3,z)$ (since $t_0$ and $t_3$ are the only transitions that can occur directly before $t_1$) and uses this as the ``incoming size'' of $z$.
	Hence, it obtains $\Size(t_1,x) = \Size(t_1,y) = z + z = 2 \cdot z$.
	The approach of \cite{koat} cannot compute finite size bounds for $(t_2,x)$, $(t_2,y)$, $(t_3,x)$, and $(t_3,y)$, since it needs a runtime bound for $t_2$ to over-approximate how often the ``previous'' transition $t_2$ may have been executed.
	In contrast, our results from \Cref{sec:mrf} will enable the computation of finite size bounds for all result variables of this program, see \Cref{ex:mrf_final}.
\end{example}

So size bounds on previous transitions are needed to compute runtime bounds, and similarly, runtime bounds are needed to compute size bounds.
The algorithm for the computation of size bounds in \cite{koat} is not needed to
understand the techniques presented in the current paper and thus, we use it as a
black \pagebreak box.

\section{Runtime Bounds by Multiphase Ranking Functions}
\label{sec:mrf}
The approach for computing runtime bounds in \cite{koat} relies on polynomial ranking functions (see, e.g., \cite{podelski2004complete,bradley_polyranking}).
In this section, we extend this approach to so-called multiphase-linear ranking functions ($\MRFs$) (see, e.g., \cite{genaim_mrf,leike_linear,genaim_mrf_recurrent,dblp:journals/sttt/yuanls21}).
Our experiments in \Cref{sec:eval} demonstrate that this improves its power significantly.

In \cite{genaim_mrf} it was already shown how to obtain a runtime bound from an $\MRF$ for a full integer program.
We now adapt this result to our modular approach which allows for the computation of $\MRFs$ for parts of the program (\cref{thm:time_bound}).

\subsection{Multiphase-Linear Ranking Functions}

As mentioned, the idea of ranking functions is to construct a function which decreases by at least one in every evaluation step when a specific transition is applied.
Moreover, the ranking function has to be non-negative before we apply a transition.
Thus, if the function becomes negative, then the program terminates.

An $\MRF$ extends this idea and uses a ranking function $f_i$ for every ``phase'' $1 \leq i \leq d$ of a program.
When the phases $1$ to $i - 1$ are finished,
the functions $f_1,\ldots,f_{i-1}$ remain negative and decreasing, but now the function $f_i$ becomes decreasing as well.
If all functions are negative, then the program terminates.

\cref{def:mrf} corresponds to so-called nested $\MRFs$ from \cite{genaim_mrf,leike_linear}.
Here, the sum of\linebreak
$f_{i-1}$ and $f_i$ must be larger than the updated function $f_i$ for all $i$.
We set $f_0$ to 0.
Then $f_0 + f_1 = f_1$ must be decreasing with each update.
If $f_1$ becomes negative, then $f_1 + f_2 < f_2$ and thus, $f_2$ has to be decreasing with every update, and so on until $f_d$ becomes decreasing.
The program eventually terminates, since $f_d$ must be non-negative whenever the program can be executed further.
We restrict ourselves to such ``nested'' $\MRFs$, as they are particularly easy to automate (i.e., one does not have to consider the mapping of evaluation steps to the different phases).
As usual, we use an SMT solver to search for $\MRFs$ automatically.

In contrast to \cite{genaim_mrf,leike_linear}, we define $\MRFs$ for sub-programs $\TSet'_{>} \subseteq \TSet' \subseteq \TSet$ which is crucial for our modular approach (see \cref{thm:time_bound}).
Let $\linpoly{\PVSet}$ denote the set of linear polynomials (i.e., of degree at most 1) over $\ZZ$ in the variables $\PVSet$.

\begin{definition}[$\MRFs$ for Sub-Programs]
	\label{def:mrf}
	Let $\emptyset\neq\TSet'_{>} \subseteq \TSet' \subseteq \TSet$ and $d \geq 1$.
	A tuple $f = (f_1,\dots,f_d)$ of functions $f_1,\ldots,f_d: \LSet \rightarrow \linpoly{\PVSet}$ is an \emph{$\MRF$ of depth}
	$d$ for $\TSet'_{>}$ and $\TSet'$ if for all evaluation steps $(\location,\sigma)\rightarrow_t(\location',\sigma')$:
	\begin{itemize}
		\item[(a)] If $t \in \TSet'_{>}$, then we have $\exacteval{f_{i-1}(\location)}{\sigma}
				+ \exacteval{f_i(\location)}{\sigma} \geq \exacteval{ f_i(\location')}{\sigma'} + 1$ for all $1 \leq i \leq d$ and $\exacteval{f_{d}(\location)}{\sigma} \geq 0$.
		\item[(b)] If $t \in \TSet'\setminus \TSet'_{>}$, then we have $\exacteval{f_i(\location)}{\sigma} \geq \exacteval{ f_i(\location')}{\sigma'}$ for all $1 \leq i \leq d$.
	\end{itemize}
	Here, we set $f_0(\location)\!=\!0$ for all $\location\!\in\!\LSet$.
	We say that $\TSet'\setminus\TSet'_{>}$ is the set of \emph{non-increasing}\linebreak transitions and $\TSet'_{>}$ is the set of \emph{decreasing} transitions of the $\MRF$ $f$.
\end{definition}

The definitions of $\MRFs$ and  of linear ranking functions
coincide in the special case of a single phase (i.e., if $d = 1$).
Note that for $d > 1$, the requirement for decreasing transitions in (a) does not imply
the requirement for non-increasing transitions in (b). The reason is that for decreasing
transitions, $f_i$ may increase in the beginning (if $f_{i-1}$ is large enough), \pagebreak
because eventually $f_{i-1}$ will become negative. In contrast, for non-increasing
transitions, (b) prohibits any increase of $f_i$, since
the $\MRF$ does
not represent any bound on the number of applications of these non-increasing
transitions. Thus, we cannot  replace (b) by
$\exacteval{f_{i-1}(\location)}{\sigma}
				+ \exacteval{f_i(\location)}{\sigma} \geq \exacteval{
                                f_i(\location')}{\sigma'}$, because then
such transitions might
make $f_i$ arbitrarily large if their repeated application 
does not change a positive $f_{i-1}$.

\begin{example}
	\label{ex:running_mrf}
	Consider again the integer program in \cref{fig:running} and let $\TSet'_{>}= \{t_2\}$ and $\TSet' = \{t_2, t_3\}$.
	(See \Cref{algorithm:mprf} for our heuristic to choose $\TSet'_{>}$ and $\TSet'$.)
	An execution of the loop $\TSet'_{>}= \{t_2\}$ has two phases:
	In the first phase, both $x$ and $y$ are positive.
	In every iteration, $x$ increases until $y$ is $0$.
	The second phase starts when $y$ is negative.
	This phase ends when $x$ is negative, since then the guard $x>0$ is not satisfied anymore.
	We now show that the tuple $(f_1, f_2)$ is an $\MRF$ for $\TSet'_{>} = \{t_2\}$ and
	$\TSet' = \{t_2, t_3\}$ where $f_1(\location_1) = f_1(\location_2) = y+1$ and  $f_2(\location_1) = f_2(\location_2) = x$.

	Since $t_2$ has the update function $\eta$ with $\eta(x) = x+y$ and $\eta(y) = y-1$, for any\linebreak
	evaluation step $(\location_2,\valuation)\rightarrow_{t_2}(\location_2,\valuation')$, we have $\valuation'(x) = \valuation(\eta(x)) = \valuation(x) + \valuation(y)$ and\linebreak
	$\valuation'(y) = \valuation(\eta(y)) = \valuation(y) -1$.
	Hence, $\exacteval{f_{0}(\location_2)}{\valuation}
		+ \exacteval{f_1(\location_2)}{\valuation} = 0 + \exacteval{y + 1}{\valuation}
		= \exacteval{y}{\valuation} + 1 = \exacteval{y + 1}{\valuation'} + 1 = \exacteval{f_1(\location_2)}{\valuation'} + 1$ and $\exacteval{f_{1}(\location_2)}{\valuation}
		+ \exacteval{f_2(\location_2)}{\valuation} = \exacteval{y + 1}{\valuation}
		+ \exacteval{x}{\valuation}
		= \exacteval{x}{\valuation} + \exacteval{y}{\valuation} + 1 = \exacteval{x}{\valuation'} + 1 = \exacteval{f_2(\location_2)}{\valuation'} + 1$. Moreover, due to the guard $x > 0$, $\exacteval{x>0}{\valuation} = \true$ implies $\exacteval{f_{2}(\location_2)}{\valuation} = \exacteval{x}{\valuation} \geq 0$.
	Note that neither $y+1$ (as $y$ is not bounded) nor $x$ (as $x$ might increase) are ranking functions for $t_2$.

	Similarly, since the update function $\eta$ of $t_3$ does not modify $x$ and $y$, for every evaluation step $(\location_2,\valuation)\rightarrow_{t_3}(\location_1,\valuation')$, we have $\valuation'(x) = \valuation(\eta(x)) = \valuation(x)$ and $\valuation'(y) = \valuation(\eta(y)) = \valuation(y)$.
	Hence, $\exacteval{f_1(\location_2)}{\valuation} = \exacteval{y + 1}{\valuation} = \exacteval{y}{\valuation} + 1 = \exacteval{y+1}{\valuation'} = \exacteval{f_1(\location_1)}{\valuation'}$ and $\exacteval{f_2(\location_2)}{\valuation} = \exacteval{x}{\valuation} = \exacteval{x}{\valuation'} = \exacteval{f_2(\location_1)}{\valuation'}$.
\end{example}

\subsection{Computing Runtime Bounds}
\label{Computing Runtime Bounds}
We now show how to compute runtime bounds using $\MRFs$.
As in \cite{koat}, for a sub-program $\TSet'$, the \emph{entry transitions of a location} $\location$ are all transitions outside $\TSet'$ which reach $\location$.
The \emph{entry locations of} $\TSet'$ are all locations where an evaluation of the sub-program $\TSet'$ can begin.
Finally, the \emph{entry transitions of} $\TSet'$ are all entry transitions to entry locations of $\TSet'$.
\begin{definition}[Entry Transitions and Entry Locations]
	Let $\emptyset\neq \TSet' \subseteq \TSet$.
	We define the set of \emph{entry transitions of} $\location \in \LSet$ as $\TSet_{\location} = \braced{t \mid t=(\location',\guard,\update,\location)\wedge t\in\TSet\setminus\TSet'}$.
	The set of \emph{entry locations} is $\mathcal{E}_{\TSet'}
		= \braced{\location_{in} \mid \TSet_{\location_{\mathit{in}}} \neq \emptyset \wedge \exists \location': (\location_{in}, \guard,\update, \location') \in \TSet'}$.
	Finally, the \emph{entry transitions of} $\TSet'$ are $\mathcal{ET}_{\TSet'}
		= \bigcup_{\location \in \mathcal{E}_{\TSet'}} \TSet_{\location}$.
\end{definition}

\begin{example}
	\label{ex:entry}
	Again, consider the integer program in \cref{fig:running} and $\TSet'=\{t_2, t_3\}$.
	Then we have $\TSet_{\location_2} = \{t_1\}$, $\mathcal{E}_{\TSet'}
		= \{\location_2\}$, and $\mathcal{ET}_{\TSet'} = \{t_1\}$.
\end{example}

In \cite[Lemma 6]{genaim_mrf}, the authors considered programs consisting of a single looping transition and showed that an $\MRF$ for the loop yields a linear bound on the possible number of its executions.
We now generalize their lemma to our modular setting where we regard sub-programs $\TSet'$
instead of the full program $\TSet$.\footnote{So in
the special case where $\TSet'_{>} = \TSet'$ and $\TSet'$ is a singleton, our \Cref{lem:coef}
corresponds to \cite[Lemma 6]{genaim_mrf} for nested $\MRFs$.}
The sub-program $\TSet'$ may contain arbitrary many transitions and loops.

For a start configuration $(\location,\valuation)$ where $\location$ is an entry location
of $\TSet'$ and an $\MRF$\linebreak $f = (f_1,\ldots,f_d)$ for $\TSet'_{>}$ and
$\TSet'$, \cref{lem:coef} gives a bound $\beta \in \NN$ which ensures that whenever there
is an evaluation of $\TSet'$ that begins with $(\location,\valuation)$ and where
transitions\linebreak from  $\TSet'_{>}$ are applied at least $\beta$ times, then all ranking
functions in $f$ have become negative. 
As $f$ is an $\MRF$ (and thus, in every application of a transition from $\TSet'_{>}$, some $f_i$ must be decreasing and non-negative), this implies that in any evaluation of $\TSet'$ starting in $(\location,\valuation)$, transitions from $\TSet'_{>}$ can be applied \emph{at most} $\beta$ times.
Since
the bound $\beta$ depends \emph{linearly} on the values $\exacteval{f_1(\location)}{\valuation},\dots,\exacteval{f_d(\location)}{\valuation}$ of the ranking functions in the start configuration $(\location,\valuation)$ and since all ranking functions $f_i$ are linear as well, this means that we have inferred a linear bound on the number of applications of transitions from $\TSet'_{>}$.
However, this is only a \emph{local} bound w.r.t.\ the values of the variables at the start of the sub-program $\TSet'$.
We lift these local bounds to global runtime bounds for the full program in \Cref{thm:time_bound}.
See \cref{app:proofs}
for the proofs of both \cref{lem:coef,thm:time_bound}.

\begin{lemma}[Local Runtime Bound for Sub-Program]
	\label{lem:coef}
	Let $\emptyset\neq \TSet'_{>}\subseteq\TSet' \subseteq \TSet$, $\location \in \mathcal{E}_{\TSet'}$, $\valuation \in \Valuation$, and let $f=(f_1,\ldots,f_d)$ be an $\MRF$ for $\TSet'_{>}$ and $\TSet'$.
	For all $1 \leq i \leq d$, we define the constants $\gamma_i \in \QQ$ and $\beta \in \NN$ with $\gamma_i,\beta > 0$:
	\begin{itemize}
		\item[$\bullet$] $\gamma_1 = 1$ and $\gamma_i = 2 + \tfrac{\gamma_{i-1}}{i-1} + \tfrac{1}{(i-1)!}$ for $i > 1$
		\item[$\bullet$] $\beta = 1 + d! \cdot \gamma_d \; \cdot \; \max\{0, \sigma(f_1(\location)), \ldots, \sigma(f_d(\location)) \}$
	\end{itemize}
	Then for any evaluation $(\location,\valuation) \, (\rightarrow^*_{\TSet'\setminus\TSet'_{>}} \circ \rightarrow_{\TSet'_{>}})^n \, (\location',\valuation')$ with $n \geq \beta$ and any $1 \leq i \leq d$, we have $\sigma'(f_i(\location')) < 0$.
\end{lemma}
\makeproof{lem:coef}{
	We first present lemmas which give an upper and a lower bound for sums of powers.
	These lemmas will be needed in the proof of \Cref{lem:coef}.
	\begin{lemma}[Upper Bound for Sums of Powers]
		\label{lem:upperbound}
		For any $i\geq 2$ and $k \geq 1$ we have $\sum_{j=1}^{k-1} j^{i-2} \leq \tfrac{k^{i-1}}{i-1}$.
	\end{lemma}

	\begin{myproof}
		We have $\sum_{j=1}^{k-1}
			j^{i-2} \; \leq \; \sum_{j=1}^{k-1} \int_{j}^{j+1} x^{i-2} \, dx \; \leq \; \int_{0}^{k} x^{i-2} \, dx \; = \; \tfrac{k^{i-1}}{i-1}$.
	\end{myproof}

	For the lower bound, we use the summation formula of Euler (see, e.g., \cite{koenigsberger2003}).

	\begin{lemma}[Summation Formula of Euler]
		\label{lem:euler}
		We define the periodic function $H : \R \rightarrow \R$ as $H(x) =	x - \lfloor x \rfloor - \tfrac{1}{2}$ if $x \in\R\setminus\ZZ$ and as $H(x) = 0$ if $x\in\ZZ$.
		Note that $H(x)$ is bounded by $-\tfrac{1}{2}$ and $\tfrac{1}{2}$.
		Then for any continuously differentiable function $f: [1,n] \rightarrow \C$ with $n\in\NN$, we have $\sum_{j=1}^{k} f(j) \; = \; \int_{1}^{k} f(x) \, dx + \tfrac{1}{2}\cdot (f(1)+f(k)) + \int_{1}^{k} H(x)\cdot f'(x)\, dx$.
	\end{lemma}

	This then leads to the following result.

	\begin{lemma}[Lower Bound for Sums of Powers]
		\label{lem:lowerbound}
		For any $i\geq 2$ and $k \geq 1$ we have $\sum_{j=1}^{k-1} j^{i-1}\geq \tfrac{k^{i}}{i} - k^{i-1}$.
	\end{lemma}

	\begin{myproof}
		Consider $f(x) = x^i$ with the derivative $f'(x) = i \cdot x^{i-1}$.
		We get
		\begin{align*}
			        & \sum\nolimits_{j=1}^{k} j^i                                                                                                  \\
			{} = {} & \int_{1}^{k} x^i \,dx + \tfrac{1}{2} \cdot (1+k^i) + \int_{1}^{k} H(x) \cdot i \cdot x^{i-1} \, dx \tag{by \Cref{lem:euler}} \\
			{} = {} & \tfrac{k^{i+1}}{i+1} - \tfrac{1}{i+1}
			+ \tfrac{1}{2} \cdot (1+k^i) + \int_{1}^{k} H(x) \cdot i \cdot x^{i-1} \, dx                                                           \\
			{} = {} & \tfrac{k^{i+1}}{i+1}
			+ R \tag{for $R = - \tfrac{1}{i+1}
					+ \tfrac{1}{2} \cdot (1+k^i) + \int_{1}^{k} H(x) \cdot i \cdot x^{i-1} \, dx$}
		\end{align*}

		Since $\abs{H(x)} \leq \tfrac{1}{2}$, we have $\abs{\int_{1}^{k} H(x) \cdot i \cdot x^{i-1} \, dx}
			\; \leq \; \tfrac{1}{2} \cdot \abs{\int_{1}^{k} i \cdot x^{i-1} \, dx} \; =\; \tfrac{1}{2} \cdot i \cdot \abs{\tfrac{k^i}{i} - \tfrac{1}{i}}
			\; = \; \tfrac{k^i - 1}{2}$. Thus, we obtain
		\[
			- \tfrac{1}{i+1} + \tfrac{1}{2} \cdot (1+k^i) + \tfrac{k^i - 1}{2} \geq R \geq - \tfrac{1}{i+1} + \tfrac{1}{2} \cdot (1+k^i) - \tfrac{k^i - 1}{2}
		\]
		or, equivalently $- \tfrac{1}{i+1} + k^i \geq R \geq - \tfrac{1}{i+1} +1$.
		This implies $k^i > R > 0$.	Hence, we get $\sum_{j=1}^{k} j^i \; = \; \tfrac{k^{i+1}}{i+1}
			+ R \; \geq \; \tfrac{k^{i+1}}{i+1}$ and thus, $\sum_{j=1}^{k-1} j^i \; = \; \sum_{j=1}^{k} j^i - k^i \; \geq \; \tfrac{k^{i+1}}{i+1}
			- k^i$. With the index shift $i\rightarrow i - 1$ we finally obtain the lower bound $\sum_{j=1}^{k-1}
			j^{i-1}\geq \tfrac{k^{i}}{i} - k^{i-1}$.
	\end{myproof}

	\noindent
	\emph{Proof of \Cref{lem:coef}.}
	To ease notation, in this proof $\location_0$ does not denote the initial location of the program $\TSet$, but an arbitrary location from $\LSet$.
	Then we can write $(\location_0,\valuation_0)$ instead of $(\location,\valuation)$, $(\location_n,\valuation_n)$ instead of $(\location',\valuation')$, and consider an evaluation
	\[
		(\location_{0},\valuation_{0}) \, (\rightarrow^*_{\TSet'\setminus\TSet'_{>}} \circ \rightarrow_{\TSet'_{>}}) \, (\location_{1},\valuation_{1}) \, (\rightarrow^*_{\TSet'\setminus\TSet'_{>}} \circ \rightarrow_{\TSet'_{>}}) \, \ldots \, (\rightarrow^*_{\TSet'\setminus\TSet'_{>}} \circ \rightarrow_{\TSet'_{>}}) \, (\location_{n},\valuation_{n}).
	\]
	Let $M = \max\{0,\exacteval{f_1(\location_0)}{\valuation_0}, \dots, \exacteval{f_d(\location_0)}{\valuation_0}\}$.
We first prove that for all $1 \leq i \leq d$ and all $0 \leq k \leq n$, we have
\begin{equation}
        \label{Lemma18aux}
\sigma_k(f_i(\location_k)) \; \leq \; -k  \text{ if M = 0} \quad \text{and} \quad
\sigma_k(f_i(\location_k)) \; \leq \; \gamma_i \cdot M \cdot k^{i-1}
	- \tfrac{k^i}{i!} \; \text{ if $M > 0$.}
	\end{equation}

	The proof is done by induction on $i$.
	So in the base case, we have $i = 1$.
	Since $\gamma_1 = 1$, we have to show that $\exacteval{ f_1(\location_k)}{\sigma_k} \leq M \cdot k^{0} - \tfrac{k^1}{1!} = M - k$.

For all
$0 \leq j \leq k-1$,
the step from $(\location_j,\valuation_j)$ to $(\location_{j+1},\valuation_{j+1})$
 corresponds to the evaluation of transitions from $\TSet'\setminus \TSet'_{>}$
followed by 
a transition from $\TSet'_{>}$,
i.e., we have
$(\location_j,\valuation_j) \rightarrow^*_{\TSet'\setminus\TSet'_{>}}
 (\location'_j,\valuation'_j)  \rightarrow_{\TSet'_{>}}
 (\location_{j+1},\valuation_{j+1})$ for some configuration
 $(\location'_j,\valuation'_j)$. Since $f$ is an $\MRF$ and all transitions in  
$\TSet'\setminus\TSet'_{>}$ are non-increasing, we obtain
 $\valuation_j(f_1(\location_j)) \geq \valuation'_{j}(f_1(\location'_{j}))$.
Moreover, since the transitions in $\TSet'_{>}$ are decreasing, we have
$\valuation'_j(f_{0}(\location'_j))
 + \valuation'_j(f_{1}(\location'_j))
= \valuation'_j(f_{1}(\location'_j))
\geq \valuation_{j+1}(f_{1}(\location_{j+1})) + 1$.
So together, this implies 
$\valuation_j(f_{1}(\location_j)) \geq \valuation_{j+1}(f_{1}(\location_{j+1})) + 1$ and
thus, 
$\exacteval{f_1(\location_0)}{\sigma_0}\geq \exacteval{ f_1(\location_1)}{\sigma_1} + 1\geq\ldots\geq \exacteval{ f_1(\location_k)}{\sigma_k} + k$ or equivalently, $\exacteval{f_1(\location_0)}{\sigma_0} - k\geq \exacteval{ f_1(\location_k)}{\sigma_k}$.
	Furthermore, we have $\exacteval{f_1(\location_0)}{\sigma_0}\leq \max\{0,\exacteval{f_1(\location_0)}{\valuation_0},\dots,\exacteval{f_d(\location_0)}{\valuation_0}\} = M$.
	Hence, we obtain $\exacteval{
			f_1(\location_k)}{\sigma_k} \leq \exacteval{f_1(\location_0)}{\sigma_0}
		- k \leq M - k$. So in particular, if $M = 0$, then we have $\sigma_k(f_1(\location_k)) \leq -k$.

	In the induction step, we assume that
for all $0 \leq k \leq n$, we have $\sigma_k(f_{i-1}(\location_k)) \, \leq \,
-k$ if $M = 0$ and 
$\sigma_k(f_{i-1}(\location_k)) \; \leq \; \gamma_{i-1} \cdot M \cdot k^{i-2}
- \tfrac{k^{i-1}}{(i-1)!}$ if $M > 0$.
	To show that the inequations also hold for $i$, we first transform $\sigma_k(f_{i}(\location_k))$ into a telescoping sum.
	\begin{align}
		\exacteval{f_i(\location_k)}{\sigma_k} = \exacteval{f_i(\location_0)}{\sigma_0} + \sum_{j=0}^{k-1} (\exacteval{f_i(\location_{j+1})}{\sigma_{j+1}} - \exacteval{f_i(\location_{j})}{\sigma_{j}}) \nonumber
	\end{align}

For all
$0 \leq j \leq k-1$,
the step from $(\location_j,\valuation_j)$ to $(\location_{j+1},\valuation_{j+1})$ again
has the form
$(\location_j,\valuation_j) \rightarrow^*_{\TSet'\setminus\TSet'_{>}}
 (\location'_j,\valuation'_j)  \rightarrow_{\TSet'_{>}}
 (\location_{j+1},\valuation_{j+1})$ for some configuration
 $(\location'_j,\valuation'_j)$. Since $f$ is an $\MRF$ and all transitions in  
$\TSet'\setminus\TSet'_{>}$ are non-increasing, we obtain
 $\valuation_j(f_{i-1}(\location_j)) \geq \valuation'_{j}(f_{i-1}(\location'_{j}))$
and $\valuation_j(f_i(\location_j)) \geq \valuation'_{j}(f_i(\location'_{j}))$.
Moreover, since the transitions in $\TSet'_{>}$ are decreasing, we have
$\valuation'_j(f_{i-1}(\location'_j))
 + \valuation'_j(f_{i}(\location'_j)) \geq \valuation_{j+1}(f_{i}(\location_{j+1})) + 1$.
 So together, this implies 
$\valuation_j(f_{i-1}(\location_j))
 + \valuation_j(f_{i}(\location_j)) \geq \valuation_{j+1}(f_{i}(\location_{j+1})) + 1$
        or equivalently,
$\exacteval{f_i (\location_{j+1})}{\sigma_{j + 1}} - \exacteval{f_i
(\location_j)}{\sigma_{j}} < \exacteval{f_{i-1} (\location_{j})}{\sigma_{j}}$.
	Hence, we obtain
	\begin{align*}
            \exacteval{f_i(\location_{k})}{\sigma_{k}} &= \exacteval{f_i(\location_{0})}{\sigma_{0}} + \sum_{j=0}^{k-1}
		(\exacteval{f_i(\location_{j+1})}{\sigma_{j+1}}
		- \exacteval{f_i(\location_{j})}{\sigma_{j}})                       \\
                                                           &< \exacteval{f_i(\location_{0})}{\sigma_{0}} + \sum_{j=0}^{k-1}
		\exacteval{f_{i-1}(\location_{j})}{\sigma_{j}}.
	\end{align*}

	If $M = 0$, then we obviously have $\sigma_{0}(f_i (\location_{0})) \leq 0$ for all $1\leq i \leq d$.
	For $k \geq 1$, we obtain
	\begin{align*}
& \exacteval{f_i(\location_0)}{\sigma_0} + \sum_{j=0}^{k-1}
		\exacteval{f_{i-1}(\location_{j})}{\sigma_{j}}\\
                {} \leq {} & 0 +  \sum_{j=0}^{k-1} -j \tag{ by the induction hypothesis}
		\\
                {} \leq {} & - k + 1.
                \end{align*}
Hence, we have $\exacteval{f_i(\location_k)}{\sigma_k} <  - k + 1$ and thus,
$\exacteval{f_i(\location_k)}{\sigma_k} \leq -k$.

If $M > 0$, then we obtain 
\begin{align*}
& \exacteval{f_i(\location_0)}{\sigma_0} + \sum_{j=0}^{k-1}
		\exacteval{f_{i-1}(\location_{j})}{\sigma_{j}}\\
		{} \leq {} & 2 \cdot M + \sum_{j=1}^{k-1}
		\exacteval{f_{i-1}(\location_{j})}{\sigma_{j}}
		\tag{ as $\exacteval{f_i(\location_0)}{\sigma_0} \leq M$ and $\exacteval{f_{i-1}(\location_{0})}{\sigma_{0}} \leq M$
		}
		\\
		{} \leq {} & 2 \cdot M + \sum_{j=1}^{k-1}
		( \gamma_{i-1} \cdot M \cdot j^{i-2} - \tfrac{j^{i-1}}{(i-1)!}) \tag{ by the induction hypothesis}
		\\
		{} = {}    & 2 \cdot M + \gamma_{i-1} \cdot M \cdot \left(\sum_{j=1}^{k-1}
		j^{i-2}\right) - \tfrac{1}{(i-1)!} \cdot \left(\sum_{j=1}^{k-1}
		j^{i-1}\right)                                                                                                                                                                                 \\
		{} \leq {} & 2 \cdot M + \gamma_{i-1} \cdot M \cdot \tfrac{k^{i-1}}{i-1} - \tfrac{1}{(i-1)!} \cdot \left(\tfrac{k^i}{i} - k^{i-1} \right) \tag{by \cref{lem:upperbound,lem:lowerbound}} \\
		{} = {}    & 2 \cdot M + \gamma_{i-1} \cdot M \cdot \tfrac{k^{i-1}}{i-1}
		+ \tfrac{k^{i-1}}{(i-1)!} - \tfrac{k^i}{i!}                                                                                                                                             \\
		{} \leq {} & 2 \cdot M \cdot k^{i-1}+ \gamma_{i-1} \cdot M \cdot \tfrac{k^{i-1}}{i-1}
		+\tfrac{k^{i-1}}{(i-1)!} - \tfrac{k^i}{i!}                                                                                                                                              \\
		{} \leq {} & M \cdot k^{i-1} \cdot \left( \underbrace{2 + \tfrac{\gamma_{i-1}}{i-1}
		+ \tfrac{1}{(i-1)!}}_{\gamma_i} \right) - \tfrac{k^i}{i!} \tag{as $M\geq 1$}                                                                                                  \\
		{} = {}    & M \cdot k^{i-1} \cdot \gamma_i -\tfrac{k^i}{i!}.
	\end{align*}
        Hence, \eqref{Lemma18aux} is proved.

In the case $M = 0$, \eqref{Lemma18aux} implies $\sigma_n(f_i(\location_n)) \leq
-n \leq -\beta = -1 < 0$ for all $1\leq i \leq d$ which proves the lemma.

	Hence, it remains to regard the case $M > 0$.
	Now \eqref{Lemma18aux} implies
	\begin{equation}
		\label{first step}
		\sigma_n(f_i(\location_n)) \; \leq \; \gamma_i \cdot M \cdot n^{i-1} - \tfrac{n^i}{i!}.
	\end{equation}

	We now prove that for $i > 1$ we always have $i! \cdot \gamma_i \geq (i-1)! \cdot \gamma_{i-1}$.
	\begin{align*}
		           & i! \cdot \gamma_i                                                       \\
		{} = {}    & i! \cdot \left(2 + \tfrac{\gamma_{i-1}}{i-1} + \tfrac{1}{(i-1)!}\right) \\
		{} = {}    & i! \cdot 2 + i \cdot (i-2)! \cdot \gamma_{i-1} + i                      \\
		{} \geq {} & (i-1) \cdot (i-2)! \cdot \gamma_{i-1}                                   \\
		{} = {}    & (i-1)! \cdot \gamma_{i-1}.
	\end{align*}
	Thus,
	\begin{equation}
		\label{second step}
		d! \cdot \gamma_d \geq i! \cdot \gamma_i \quad \text{for all } 1 \leq i \leq d.
	\end{equation}

	Hence, for $n \geq \beta = 1 + d! \cdot \gamma_d \cdot M$ we obtain:
	\begin{align*}
		           & \sigma_n(f_i(\location_n))                                            \\
		{} \leq {} & \gamma_i \cdot M \cdot n^{i-1} - \tfrac{n^i}{i!} \tag{
		by \eqref{first step}}                                                             \\
		{} = {}    & \tfrac{n^{i-1}}{i!} \cdot \left( i! \cdot \gamma_i \cdot M - n\right) \\
		{} \leq {} & \tfrac{n^{i-1}}{i!} \cdot (\beta -1- n) \tag{
		by \eqref{second step}}                                                            \\
		{} < {}    & 0 \tag{ since $n \geq \beta$}
	\end{align*}

	Finally, to show that $\beta \in \NN$, note that by induction on $i$, one can easily prove that $(i-1)! \cdot \gamma_i \in \NN$ holds for all $i \geq 1$.
	Hence, in contrast to $\gamma_i$, the number $i! \cdot \gamma_i$ is a natural number for all $i \in \NN$.
	This implies $\beta \in \NN$.
	\qed
}

Note that the constants $\gamma_i$ do not depend on the program or the $\MRF$, and the factor $d! \cdot \gamma_d$ only depends on the depth $d$.

\begin{example}
	\label{ex:coefmrf}
	Reconsider the $\MRF = (f_1,f_2)$ that we found for $\TSet'_{>} = \{t_2\}$ and $\TSet' = \{ t_2, t_3 \}$ in \cref{ex:running_mrf}.
	The constants of \cref{lem:coef}
	are $\gamma_1= 1$ and $\gamma_2 = 2 + \tfrac{1}{1} + \tfrac{1}{1} = 4$.
	Thus, when $\TSet'$ is interpreted as a standalone program, then transition $t_2$ can be executed at most $\beta = 1 + 2! \cdot \gamma_2 \cdot \max\{0,\exacteval{f_1(\location_2)}{\valuation},\exacteval{f_2(\location_2)}{\valuation}\}
		= 1 + 8\cdot \maximum{0, \exacteval{y + 1}{\valuation},\exacteval{x}{\valuation}}$ many times when starting in $\valuation \in \Valuation$.
\end{example}

\Cref{lem:coef} yields the runtime bound
\begin{equation}
	\label{localOriginalBound}
	1 + d! \cdot \gamma_d \; \cdot \; \max\{0, f_1(\location), \ldots, f_d(\location) \}
\end{equation}
for the transitions $\TSet'_{>}$ in the standalone program consisting of the transitions $\TSet'$.
However, \eqref{localOriginalBound} is not yet a bound from $\BoundSet$, because it contains ``$\max$'' and because the polynomials $f_i(\location)$ may have negative coefficients.
To transform polynomials into (weakly monotonically increasing) bounds, we replace their coefficients by their absolute values (and denote this transformation by $\overapprox{\cdot}$).
So for example we have $\overapprox{-x+2} = \abs{-1} \cdot x + \abs{2} = x + 2$.
Moreover, to remove ``$\max$'', we replace it by addition.
In this way, we obtain the bound
\[
	\beta_{\location} \; = \; 1 + d! \cdot \gamma_d \; \cdot \; (\overapprox{f_1(\location)} + \ldots + \overapprox{f_d(\location)}).
\]

In an evaluation of the full program, we enter a sub-program $\TSet'$ by an entry transition $t\in\TSet_{\location}$ to an entry location $\location \in \mathcal{E}_{\TSet'}$.
As explained in \Cref{sec:preliminaries}, to lift the local runtime bound
$\beta_{\location}$ for $\TSet'_{>}$ to a global bound, we have to instantiate the variables in
$\beta_{\location}$ by \pagebreak (over-approximations of) the values that the variables have when reaching the sub-program $\TSet'$, i.e., after the transition $t$.
To this end, we use the size bound $\Size(t,v)$ which over-approximates the largest absolute value of $v$ after the transition $t$.
We also use the shorthand notation $\Size(t,\cdot):\PVSet \to \BoundSet$, where $\Size(t,\cdot)(v)$ is defined to be $\Size(t,v)$ and for every arithmetic expression $b$, $\Size(t,\cdot)(b)$ results from $b$ by replacing each variable $v$ in $b$ by $\Size(t,v)$.
Hence, $\Size(t,\cdot)(\beta_\location)$ is a (global) bound on the number of applications of transitions from $\TSet'_{>}$ if $\TSet'$ is entered once via the entry transition $t$.
Here, weak monotonic increase of $\beta_\location$ ensures that the over-approximation of the variables $v$ in $\beta_\location$ by $\Size(t,v)$ indeed leads to an over-approximation of $\TSet'_{>}$'s runtime.

However, for every entry transition $t$ we also have to take into account how often the sub-program $\TSet'$ may be entered via $t$.
We can over-approximate this value by $\Time(t)$.
This leads to \cref{thm:time_bound} which generalizes a result from \cite{koat} to $\MRFs$.
The analysis starts with a runtime bound $\Time$ and a size bound $\Size$ which map all
transitions resp.\ result variables to $\omega$, except for the transitions $t$ which do
not occur in cycles of $\TSet$, where $\Time(t) = 1$.
Afterwards, $\Time$ and $\Size$ are refined repeatedly.
Instead of using a single ranking function for the refinement of $\Time$ as in \cite{koat}, \Cref{thm:time_bound} now allows us to replace $\Time$ by a refined bound $\Time'$ based on an $\MRF$.

\begin{theorem}[Refining Runtime Bounds Based on $\MRFs$]
	\label{thm:time_bound}
	Let $\Time$ be a runtime bound, $\Size$ a size bound, and $\emptyset\neq \TSet'_{>} \subseteq \TSet' \subseteq \TSet$ such that $\TSet'$ does not contain any initial transitions.
	Let $f = (f_1,\ldots,f_d)$ be an $\MRF$ for $\TSet'_{>}$ and $\TSet'$.
	For any entry location $\location \in\mathcal{E}_{\TSet'}$ we define $\beta_\location = 1 + d! \cdot \gamma_d \; \cdot \; (\overapprox{f_1(\location)} + \ldots + \overapprox{f_d(\location)})$, where $\gamma_d$ is as in \cref{lem:coef}.
	Then $\Time'$ is also a runtime bound, where we define $\Time'$ by $\Time'(t) =\Time(t)$ for all $t \notin \TSet'_{>}$ and
	\[
		\Time'\left(t_{>}\right)
	= \sum\nolimits_{\location \in \mathcal{E}_{\TSet'}} \, \sum\nolimits_{t \in \TSet_\location} \, \Time(t) \cdot \exacteval{\beta_\location}{\Size(t,\cdot)} \quad \text{for all $t_>\in\TSet'_{>}$.}
	\]
        \end{theorem}
\makeproof{thm:time_bound}{
	\begin{myproof}
		We prove \cref{thm:time_bound} by showing that for all $t\in\TSet$ and all $\valuation_0 \in \Valuation$ we have
		\begin{align}
			\upeval{\Time'(t)}{\valuation_0} \geq \timeboundterm.\label{eq:sound_timebound}
		\end{align}
		The case $t \notin \TSet'_{>}$ is trivial, since $\Time'(t) = \Time(t)$ and $\Time$ is a runtime bound.

		Now we prove \eqref{eq:sound_timebound} for a transition $t_>\in\TSet'_{>}$, i.e., we show that for all $\valuation_0 \in \Valuation$ we have
		\[
			\begin{array}{r@{\;\;}c@{\;\;}l}
				\upeval{\Time'(t_>)}{\valuation_0} & =    & \sum_{\location \in \mathcal{E}_{\TSet'}} \sum_{t \in \TSet_\location} \upeval{\Time(t)}{\valuation_0} \cdot \upeval{\Size(t, \cdot)(\beta_{\location})}{\valuation_0} \\
				                                   & \geq & \timeboundterm[t_>].
			\end{array}
		\]
		So let $(\location_0, \valuation_0) \, (\rightarrow^* \circ \rightarrow_{t_>})^k \, (\location, \valuation)$ and we have to show $\upeval{\Time'(t_>)}{\valuation_0} \geq k$.
		If $k = 0$, then we clearly have $\upeval{\Time'\left(t_{>}\right)}{\valuation_0} \geq 0 = k$.
		Hence, we consider $k > 0$.
		We represent the evaluation as follows:
		\[
			\begin{array}{l@{\quad}c@{\quad}l@{\quad}c}
				(\actl_0, \actstate_0)         & \rightarrow^{\tilde{k}_0}_{\TSet \setminus \TSet'}     & (\prel_1, \prestate_1) & \rightarrow^{k_1'}_{\TSet'} \\
				(\actl_1, \actstate_1)         & \rightarrow^{\tilde{k}_1}_{\TSet \setminus \TSet'}     & (\prel_2, \prestate_2) & \rightarrow^{k_2'}_{\TSet'} \\
				                               &                                                        & \ldots                                               \\
				(\actl_{m-1}, \actstate_{m-1}) & \rightarrow^{\tilde{k}_{m-1}}_{\TSet \setminus \TSet'} & (\prel_m, \prestate_m) & \rightarrow^{k_m'}_{\TSet'} \\
				(\actl_m, \actstate_m)
			\end{array}
		\]
		So for the evaluation from $(\actl_i, \actstate_i)$ to $(\prel_{i+1}, \prestate_{i+1})$ we only use transitions from $\TSet\setminus\TSet'$, and for the evaluation from $(\prel_i, \prestate_i)$ to $(\actl_i, \actstate_i)$ we only use transitions from $\TSet'$.
		Thus, $t_>$ can only occur in the following finite sequences of evaluation steps:
		\begin{equation}
			\label{SubsetEvaluation}
			(\prel_i, \prestate_i) \rightarrow_{\TSet'} (\prel_{i,1}, \prestate_{i,1}) \rightarrow_{\TSet'} \dots \rightarrow_{\TSet'} (\prel_{i,k_i'-1}, \prestate_{i,k_i'-1}) \rightarrow_{\TSet'} (\actl_i, \actstate_i).
		\end{equation}
		For every $1 \leq i \leq m$, let $k_i \leq k_i'$ be the number of times that $t_>$ is used in the evaluation \eqref{SubsetEvaluation}.
		Clearly, we have
		\begin{equation}
			\label{SumDecreasingTransition}
			\sum_{i=1}^{m} k_i = k.
		\end{equation}

		By \cref{lem:coef}, all functions $f_1,\dots,f_d$ are negative after executing $t_>$ at least $1 + d! \cdot \gamma_d \cdot\max\{0, \prestate_i(f_1(\prel_i)),\dots, \prestate_i(f_d(\prel_i))\}$ times in an evaluation with $\TSet'$.
		If all the $f_i$ are negative, then $t_>$ cannot be executed anymore as
		$f$ is an $\MRF$ for $\TSet'_{>}$ with $t_> \in \TSet'_{>}$ and $\TSet'$.
		Thus, for all $1 \leq i \leq m$ we have
		\begin{equation}
			\label{BoundKi}
			1 + d! \cdot \gamma_d\cdot \max\left\{0,\exacteval{f_1(\prel_i)}{\prestate_i},\dots,\exacteval{f_d(\prel_i)}{\prestate_i}\right\} \geq k_i.
		\end{equation}
		Let $t_i$ be the entry transition reaching $(\prel_i, \prestate_i)$, i.e., $\prel_i \in \mathcal{E}_{\TSet'}$ and $t_i \in \TSet_{\prel_i}$.
		As $(\location_0, \valuation_0) \rightarrow^*_\TSet \circ \rightarrow_{t_i}
			(\prel_i, \prestate_i)$, by \cref{def:size_bound} we have $\upeval{\Size(t_i, v)}{\valuation_0} \geq |\prestate_i(v)|$ for all $v \in \PVSet$ and thus,
		\begin{align*}
			           & \upeval{\Size(t_i, \cdot)(\beta_{\prel_i})}{\valuation_0}
			\\
			{} \geq {} & \upeval{\beta_{\prel_i}}{\prestate_i} \tag{ since $\beta_{\prel_i} \in \BoundSet$}                                                 \\
			{} \geq {} & \prestate_i(\beta_{\prel_i})                                                                                                       \\
			{} \geq {} & 1 + d! \cdot \gamma_d \cdot \max\left\{0,\exacteval{f_1(\prel_i)}{\prestate_i},\dots,\exacteval{f_d(\prel_i)}{\prestate_i}\right\}
			\tag{ by definition of $\overapprox{\cdot}$ and $\beta_{\prel_i}$}                                                                              \\
			{} \geq {} & k_i \tag{ by \eqref{BoundKi}}
		\end{align*}

		In the last part of this proof we need to analyze how often such evaluations $(\prel_i, \prestate_i) \rightarrow^*_{\TSet'} (\actl_i, \actstate_i)$ can occur.
		Again, let $t_i$ be the entry transition reaching $(\prel_i, \prestate_i)$.
		Every entry transition $t_i$ can occur at most $\upeval{\Time(t_i)}{\valuation_0}$ times in the complete evaluation, as $\Time$ is a runtime bound.
		Thus, we have
		\begin{align*}
			\upeval{\Time'\left(t_{>}\right)}{\valuation_0} {} = {}
			           & \sum_{\location \in \mathcal{E}_{\TSet'}} \sum_{t \in \TSet_\location} \upeval{\Time(t)}{\valuation_0} \cdot \upeval{\Size(t, \cdot)(\beta_{\location})}{\valuation_0} \\
			{} \geq {} & \sum_{i=1}^m \upeval{\Size(t_i, \cdot)(\beta_{\prel_i})}{\valuation_0}                                                                                                 \\
			{} \geq {} & \sum_{i=1}^m k_i \tag{ as shown above}                                                                                                                                 \\
			{} = {}    & k \tag{ by \eqref{SumDecreasingTransition}}
		\end{align*}
	\end{myproof}
}

\begin{example}
	\label{ex:mrf_final}
	We use \cref{thm:time_bound} to compute a runtime bound for $t_2$ in \cref{fig:running}.
	In \cref{ex:entry}, we showed that $\mathcal{E}_{\{t_2,t_3\}} = \{\location_2\}$ and $\TSet_{\location_2} = \{t_1\}$.
	Thus, we obtain
	\[
		\Time(t_2)= \Time(t_1) \cdot \exacteval{\beta_{\location_2}}{\Size(t_1,\cdot)}.
	\]
	Using our calculations from \cref{ex:coefmrf} we have $\beta_{\location_2} = 1 + 2! \cdot \gamma_2 \; \cdot \; (\overapprox{f_1(\location_2)} + \overapprox{f_2(\location_2)}) = 1 + 8 \cdot (y + 1 + x) = 8\cdot x + 8 \cdot y + 9$.

	We use the runtime bound $\Time(t_1) = z$ and the size bounds $\Size(t_1,x)=\Size(t_1,y)=2 \cdot z$ from \cref{ex:time,ex:size} and get $\Time(t_2) = \Time(t_1) \, \cdot \, (8 \cdot \Size(t_1, x)+\linebreak
		8 \cdot \Size(t_1, y) + 9 ) = z \, \cdot \, \left(8 \cdot 2 \cdot z + 8 \cdot 2 \cdot z + 9 \right) = 32 \cdot z^2 + 9 \cdot z$.

	By \cref{thm:over_approx_rc,ex:time}, the runtime complexity of the program in \cref{fig:running} is at most $\sum_{j=0}^3 \Time(t_j) \; = \; 1 + z + 32\cdot z^2 + 9 \cdot z + z \; = \; 32\cdot z^2 + 11 \cdot z + 1$, resp.\ $\rc(\sigma_0) \leq 32\cdot \abs{\sigma_0(z)}^2 + 11 \cdot \abs{\sigma_0(z)} + 1$, i.e., the program's runtime complexity is at most quadratic in the initial absolute value of $z$.
	Thus, in contrast to \cite{koat}, we can now infer a finite bound on the runtime complexity of this program.
\end{example}

\subsection{Complete Algorithm}
\label{Complete Algorithm}

Based
 on \Cref{thm:time_bound}, in \Cref{algorithm:mprf}
we present our complete algorithm which improves the approach for complexity analysis of integer programs from \cite{koat} by using $\MRFs$ to infer runtime bounds.
As mentioned in \Cref{sec:preliminaries}, the computation of size bounds from \cite{koat}
is used as a black box. 
We just take the alternating repeated improvement of runtime and size bounds into account.
So in particular, size bounds are updated when runtime bounds have been updated (\cref{line:size,line:size2}).

First, we preprocess the program (\cref{line:preprocess}) by eliminating unreachable locations and transitions with unsatisfiable guards, and infer program invariants using the \tool{Apron} library \cite{apron}.
In addition, we remove variables which clearly do not have an impact on the termination behavior of the program.
Then, we set all runtime bounds for transitions outside of cycles to 1, and all other bounds to $\omega$ initially (\cref{line:initial}).

For the computation of an $\MRF$,
the considered subset $\TSet'$ has to be chosen heuristically.
We begin with regarding a strongly connected component\footnote{As usual, a graph is \emph{strongly connected} if there is a path from every node to every other node.
	A \emph{strongly connected component}
	is a maximal strongly connected sub-graph.}
(SCC) $\widetilde{\TSet}$ of the program graph in \cref{line:SCC}.
Then we try to generate an $\MRF$, and choose $\TSet'$ to consist of a maximal subset of $\widetilde{\TSet}$ where all transitions are 
non-increasing or decreasing (and at least one of the unbounded transitions $t_>$ is decreasing).
So for the program in \Cref{fig:running}, we would start with $\widetilde{\TSet} = \{t_1,
t_2, t_3 \}$, but when trying to generate an $\MRF$ for $\TSet'_{>} = \{ t_2 \}$, we can only
make $t_2$ decreasing and $t_3$ non-increasing.
For that reason, we then set $\TSet'$ to $\{t_2, t_3\}$.

\begin{figure}[t]
\begin{algorithm}[H]
	\SetAlgoLined
	\DontPrintSemicolon
	\KwIn{An integer program $\Program = (\PVSet,\LSet,\location_0,\TSet)$}
	Preprocess $\Program$\;\label{line:preprocess}
	Create an initial runtime bound $\Time$ and an initial size bound $\Size$ and set
	$d \gets 1$\;\label{line:initial}

\ForAll{SCCs $\widetilde{\TSet}$ without initial transitions of $\Program$ in topological
	order \label{line:SCC}}{
        	\Repeat{No runtime or size bound improved }{
			\ForAll{$t_> \in \widetilde{\TSet}$ with $\Time(t_>) = \omega$\label{forall-outerLoopStart}}{
				\Repeat{$\MRF$ was found or $d > \mdepth$}{
					Search for an $\MRF$ with depth $d$
                                        for a maximal subset 
	$\TSet' \subseteq \widetilde{\TSet}$ that has a subset
		$\TSet'_{>} \subseteq \TSet'$ with $t_> \in \TSet'_{>}$\linebreak such
		that all
		transitions in $\TSet'_{>}$ are decreasing \linebreak
                and all transitions in
		$\TSet' \setminus \TSet'_{>}$ are non-increasing
					\label{line:mprf}\;
					$d \gets d+1$\;
				}
				\If{$\MRF$ was found}{
					Update $\Time(t)$ for all $t \in\TSet'_{>}$  using \cref{thm:time_bound}\label{line:time}\;
				}

			}
			Update all size bounds for transitions in $\widetilde{\TSet}$ and reset $d\gets 1$\label{line:size}
                        }
               	Update all size bounds for outgoing transitions of $\widetilde{\TSet}$.         
                      \label{line:size2}  }
	
	\KwOut{Runtime Bound $\Time$ and Size Bound $\Size$\bigskip}
	\caption{Inferring Global Runtime and Size Bounds}\label{algorithm:mprf}        
\end{algorithm}
\vspace*{-.5cm}
\end{figure}
We treat the SCCs in topological order such that improved bounds for previous transitions are already available when handling the next SCC.
If an $\MRF$ was found, we update the runtime bound for all
$t \in \TSet'_{>}$ using \cref{thm:time_bound} (\cref{line:time}).
If we do not find any $\MRF$ of the given depth that makes
$t_>$ decreasing, we increase the depth and continue the search until we reach a fixed $\mdepth$.
We abort the computation of runtime bounds if no bound has been \pagebreak improved.
Here, we use a heuristic which compares polynomial bounds by their  degrees.

Finally, let us elaborate on the choice of $\mdepth$.
For example, if $\mdepth$ is $1$, then we just compute linear ranking functions.
If $\mdepth$ is infinity, then we cannot guarantee that our algorithm always terminates.
For certain classes of programs, it is possible to give a bound on $\mdepth$ such that if there is an $\MRF$ for the program, then there is also one of depth $\mdepth$ \cite{dblp:journals/sttt/yuanls21,genaim_mrf,ben-amram:2014}.
However, it is open whether such a bound is computable for general integer programs.
As the amount of time the SMT solver needs to find an $\MRF$ increases with the depth, we decided to use $5$ as a fixed maximal depth, which performed well in our examples.
Still, we provide the option for the user to change this bound.

\vspace*{-.1cm}
\section{Improving  Bounds by Control-Flow Refinement}
\label{sec:cfr}
Now we present another technique to improve the automated complexity analysis of integer programs, so-called \emph{control-flow refinement}.
The idea is to transform a program $\Program$ into a new program $\Program'$ which is ``easier'' to analyze.
Of course, we ensure that the runtime complexity of $\Program'$ is \emph{at least}
the runtime com\-plexity of $\Program$.
Then it is sound to analyze upper runtime bounds for
$\Program'$ instead of $\Program$.

Our approach is based on the \emph{partial evaluation} technique of \cite{domenech2019control}.
For termination analysis, \cite{domenech2019control} shows how to use partial evaluation of \emph{constrained Horn clauses}
locally on every SCC of the program graph.
But for complexity analysis, \cite{domenech2019control} only discusses \emph{global} partial evaluation as a preprocessing step for complexity analysis.
In \cref{sec-Partial Evaluation}, we formalize the partial evaluation technique of \cite{domenech2019control}
such that it operates directly on SCCs of integer programs and prove that it is sound for complexity analysis.
Afterwards, we improve its locality further in \cref{sec-Local Control-Flow Refinement}
such that partial evaluation is only applied \emph{on-demand} on those transitions of an integer program where our current runtime bounds are ``not yet good enough''.
Our experimental evaluation in \cref{sec:eval} shows that our local partial evaluation techniques of \cref{sec-Partial Evaluation} and \cref{sec-Local Control-Flow Refinement}
lead to a significantly stronger tool than when performing partial evaluation only globally as a preprocessing step.

As indicated in \cref{sec:intro}, the loop in \cref{fig:crf} can be transformed into two con\-secutive loops (see \cref{fig:unrolled}).
The first loop in \cref{fig:unrolled} covers the case $x < 0 \land y < z$ and the second one covers the case $x < 0 \land y \geq z$.
These cases correspond to the conjunction of the loop guard with the conditions of the two
branches of the \textbf{if}-\linebreak instruction.
Here, partial evaluation detects that these cases occur \emph{after} each other,\linebreak i.e., if $y \geq z$, then the case $y < z$ does not occur again afterwards.
In \cref{ex:unrolling}, we illustrate how our algorithm for partial evaluation performs this transformation.
In the refined program of \cref{fig:unrolled}, it is easy to see that the runtime complexity is at\linebreak most linear.
Thus, the original loop has at most linear runtime complexity as well.
Note that our tool \tool{KoAT} can also infer a linear bound for both programs
corresponding to \cref{fig:crf,fig:unrolled} \emph{without}  control-flow refinement.
In fact, for these examples it suffices to use just linear ranking functions, i.e.,  \cref{thm:time_bound} with
$\MRFs$ of depth $d = 1$.
Still, we illustrate partial evaluation using this small example to ease readability.
We will discuss the relationship between $\MRFs$ and partial evaluation at the end
of \cref{sec-Local Control-Flow Refinement}, where we also show examples to demonstrate
that these techniques do not subsume each other (see \cref{MPRBetterThanCFR,CFRBetterThanMPR}).

\subsection{SCC-Based Partial Evaluation}
\label{sec-Partial Evaluation}

We now formalize the partial evaluation of \cite{domenech2019control}
as an SCC-based refinement technique for integer programs in \cref{algorithm:partial_evaluation} and show its correctness for complexity analysis in \cref{thm:corr_partialeval}.
The intuitive idea of \cref{algorithm:partial_evaluation} is to refine a non-trivial\footnote{As usual, an SCC is \emph{non-trivial} if it contains at least one transition.}
SCC $\TSet_{\mathit{SCC}}$ of an integer program into multiple SCCs by considering ``abstract'' evaluations which do not operate on concrete states but on sets of states.
These sets of states are characterized by constraints, i.e., a constraint $\varphi$ stands for all states $\valuation$ with $\valuation(\varphi) = \true$.
To this end, we label every location $\location$ in the SCC by a constraint
$\varphi \in \ConstraintSet(\PVSet)$ which describes (a superset of) those states
$\valuation$ which can occur in this location.
So all reachable configurations with the location $\location$ have the form $(\location, \valuation)$ such that $\valuation(\varphi) = \true$.
We begin with labeling the entry locations of $\TSet_{\mathit{SCC}}$ by the constraint $\true$.
The constraints for the other locations in the SCC are obtained by considering how the updates of the transitions affect the constraints of their source locations and their guards.
The pairs of locations and constraints then become the new locations in the refined program.

\begin{figure}[t]
\begin{algorithm}[H]
	\caption{Partial Evaluation for an SCC}\label{algorithm:partial_evaluation}	\SetAlgoLined
	\DontPrintSemicolon
	\KwIn{A program $\Program = (\PVSet,\LSet,\location_0,\TSet)$ and
		a non-trivial SCC $\TSet_{\mathit{SCC}}\subseteq\TSet$ }
	$\LSet_1 \gets \{\langle \location',\true\rangle \mid \location'\in\mathcal{E}_{\TSet_{\mathit{SCC}}}\}$\label{line:init}\;
	$\TSet_{\mathit{res}} \gets \{(\location,\guard,\update, \langle\location',\true\rangle) \mid \location'\in\mathcal{E}_{\TSet_{\mathit{SCC}}}\land (\location,\guard,\update,\location') \in\TSet\setminus\TSet_\mathit{SCC}\}$\label{line:update-entry}\;
	$\LSet_0 \gets \emptyset$, $\LSet_{\mathit{done}} \gets \emptyset$\;
	\Repeat{$\LSet_1 = \emptyset$\label{line:until}}{\label{line:repeat}
		$\LSet_0 \gets \LSet_1$,
		$\LSet_1 \gets \emptyset$\label{line:L0L1}\;
		\ForAll{$\langle \location, \varphi \rangle\in \LSet_0$\label{line:foreach1}}{
			\ForAll{$(\location,\guard,\update,\location')\in\TSet_{\mathit{SCC}}$\label{line:foreach2}}{Compute
				$\varphi_{\mathit{new}}$ from $\varphi$, $\guard$, and $\update$
				such that $\models (\varphi \land \tau) \to \update(\varphi_{\mathit{new}})$
				\label{line:refine}\;
				\If{$\langle \location', \alpha_{\location'}(\varphi_{\mathit{new}})\rangle\notin \LSet_{\mathit{done}}$\label{line:ifcond}}{
					$\LSet_1 \gets \LSet_1 \cup \{\langle \location', \alpha_{\location'}(\varphi_{\mathit{new}})\rangle\}$\label{line:addto}\;
				}
				$\TSet_\mathit{res} \gets \TSet_\mathit{res} \cup \{\left(\langle \location, \varphi \rangle,\varphi\land\guard,\update, \langle \location', \alpha_{\location'}(\varphi_{\mathit{new}})\rangle\right)\}$\label{line:redirect}\;
			}
			\ForAll{$(\location,\guard,\update,\location')\in\TSet\setminus\TSet_{\mathit{SCC}}$\label{line:foreach3}}{
				$\TSet_\mathit{res} \gets \TSet_\mathit{res} \cup \left\{\left(\langle \location, \varphi \rangle,\varphi\land\guard,\update, \location'\right)\right\}$\label{line:redirect-entry}\;
			}
			$\LSet_{\mathit{done}} \gets \LSet_{\mathit{done}} \cup \{\langle \location, \varphi \rangle\}$\label{line:mark}\;
		}
	}
	\KwOut{$\Program' = (\PVSet,
			(\LSet\setminus\{\location \mid \location$ occurs as source or target in
		$\TSet_{\mathit{SCC}}\})\cup\LSet_\mathit{done}$, $\location_0,$
		$(\TSet\setminus\{(\location,\guard,\update,\location') \mid \location$ or
		$\location'$ occurs as source or target in $\TSet_{\mathit{SCC}}\}) \cup \TSet_\mathit{res})$}\vspace*{.3cm}
\end{algorithm}
\vspace*{-.3cm}
\end{figure}

Since locations can be reached by different paths, the same location may get different constraints, i.e., partial evaluation can transform a former location $\location$ into several new locations $\langle \location, \varphi_1 \rangle, \ldots, \langle \location, \varphi_n \rangle$.
So the constraints are not necessarily invariants that hold for \emph{all} evaluations that reach a location $\location$ but instead of ``widening'' (or ``generalizing'') constraints when a location can be reached by different states, we perform a case analysis and split up a location $\location$ according to the different sets of states that may reach $\location$.

After labeling every entry location $\location$ of $\TSet_{\mathit{SCC}}$ by the constraint $\true$ in \cref{line:init}
of \cref{algorithm:partial_evaluation}, we modify the entry transitions to $\location$
such that they now reach the new location $\langle \location, \true \rangle$ instead
(\cref{line:update-entry}). 
The sets $\LSet_0$ and $\LSet_1$ (the new locations whose outgoing transitions need to be processed) and $\LSet_{\mathit{done}}$ (the new locations whose outgoing transitions were already processed) are used for bookkeeping.
We then apply partial evaluation in \crefrange{line:repeat}{line:until} until there are no new locations with transitions to be processed anymore (see \cref{line:until}).

In each iteration of the outer loop in \cref{line:repeat}, the transitions of the current new locations in $\LSet_1$ are processed.
To this end, $\LSet_0$ is set to $\LSet_1$ and $\LSet_1$ is set to $\varnothing$ in \cref{line:L0L1}.
During the handling of the locations in $\LSet_0$, we might create new locations and these will be stored in $\LSet_1$ again.

We handle all locations $\langle \location, \varphi \rangle$ in $\LSet_0$ (\cref{line:foreach1}) by using all outgoing transitions $(\location,\guard,\update,\location')$.
We first consider those transitions which are part of the considered SCC (\cref{line:foreach2}), whereas the transitions which leave the SCC are handled in \cref{line:foreach3}.

The actual partial evaluation step is in \cref{line:refine}.
Given a  new location $\langle \location, \varphi \rangle$ and a transition
$t=(\location,\guard,\update,\location')$, we compute a constraint
$\varphi_{\mathit{new}}$ which over-approximates\linebreak
the set of states that can result from those states that satisfy the constraint $\varphi$ and the guard $\guard$ of the transition when applying the update $\update$.
More precisely, $\varphi_{\mathit{new}}$ has to satisfy $\models (\varphi \land \tau) \to \update(\varphi_{\mathit{new}})$, i.e., $(\varphi \land \tau) \to \update(\varphi_{\mathit{new}})$ is a tautology.
For example, if $\varphi = (x = 0)$, $\guard= \true$, and $\update(x) = x-1$, we derive $\varphi_{\mathit{new}} = (x = -1)$.
However, if we now created the new location $\langle \location', \varphi_{\mathit{new}} \rangle$, this might lead to non-termination of our algorithm.
The reason is that if $\location'$ is within a loop, then whenever one reaches $\location'$ again, one might obtain a new constraint.
In this way, one would create infinitely many new locations $\langle \location', \varphi_1\rangle, \langle \location', \varphi_2\rangle, \ldots$.
For instance, if in our example the transition with the update $\update(x) = x-1$ is a self-loop, then we would derive further new locations with the constraints $x = -2$, $x = -3$, etc.

To ensure that every former location $\location'$ only gives rise to finitely many new locations $\langle \location', \varphi \rangle$, we perform \emph{property-based abstraction} as in \cite{gallagher2019polyvariant,domenech2019control}:
For every location $\location'$ we use a finite so-called abstraction layer $\alpha_{\location'} \subseteq \{ e_1 \leq e_2 \mid e_1, e_2 \in \ZZ[\PVSet]\}$.
So $\alpha_{\location'}$ is a finite set of atomic constraints (i.e., of polynomial inequations).
Then $\alpha_{\location'}$ is extended to a function on constraints such that $\alpha_{\location'}(\varphi_{\mathit{new}})=\varphi'_{new}$ where $\varphi'_{new}$ is a conjunction of inequations from $\alpha_{\location'}$ and $\models \varphi_{\mathit{new}} \to \varphi'_{new}$.
This guarantees that partial evaluation terminates, but it can lead to an exponential
blow-up, since for every location $\location'$ there can now be
$2^{|\alpha_{\location'}|}$ many possible constraints.
In our example, instead of the infinitely many inequations $x = 0, x=-1, x=-2, \ldots$ the abstraction layer might just contain the inequation $x \leq 0$.
Then we would only obtain the new location with the constraint $x \leq 0$.

Afterwards,  in \cref{line:ifcond,line:addto}
we add the new location $\langle \location', \alpha_{\location'}(\varphi_{\mathit{new}}) \rangle$ to $\LSet_1$ if it was not processed before.
Moreover, the transition $(\location,\guard,\update,\location')$ which we used for the refinement must now get the new location as its target (\cref{line:redirect}) and $\langle \location, \varphi \rangle$ as its source.
In addition, we extend the transition's guard $\guard$ by $\varphi$.

Finally, we also have to process the transitions $(\location,\guard,\update,\location')$ which leave the SCC.
Thus, we replace the source transition $\location$ by $\langle \location, \varphi \rangle$ and again extend the guard $\guard$ of the transition by the constraint $\varphi$ in \cref{line:foreach3,line:redirect-entry}.
Since we have now processed all outgoing transitions of $\langle \location, \varphi \rangle$ we can add it to $\LSet_{\mathit{done}}$ in \cref{line:mark}.

In the end, we output the program where the considered SCC and all transitions in or out of this SCC were refined  (and thus, have to be removed from the original program).
We now illustrate \cref{algorithm:partial_evaluation} \vspace*{-.2cm} \pagebreak using the program
from \cref{fig:crf}.
\begin{wrapfigure}[12]{r}{7.4cm}
	\vspace*{-.2cm}
	\begin{minipage}{0.5\linewidth}
		\begin{tikzpicture}[->,>=stealth',shorten >=1pt,auto,node distance=4cm,semithick,initial text=$ $]
			\node[state,initial] (q1) {$\location_0$}; \node[state] (q2) [right = 1cm of q1]{$\location_1$}; \node[state] (q3) [right of=q2]{$\location_2$}; \node[state] (q4) [below = 1cm of q2]{$\location_3$}; \draw (q1) edge[above] node [align=center] {$t_0$} (q2) (q2) edge node [align=center] {$t_1: $ $\guard = x < 0$} (q3) (q3) edge[bend left] node [below,align=center] {$t_2: $ $\guard = y < z$\\
					$\update(y) = y - x$} (q2) (q3) edge[bend right] node [above,align=center] {$t_3: $ $\guard = y \geq z$\\
					$\update(x) = x + 1$} (q2) (q2) edge node [left,align=center] {$t_4: $ $\guard = x \geq 0$} (q4); \vspace*{-14cm}
		\end{tikzpicture}
	\end{minipage}
	\vspace*{-.2cm}	\caption{Integer Program Corresponding to \cref{fig:crf}}\label{fig:cfr-running}
\end{wrapfigure}
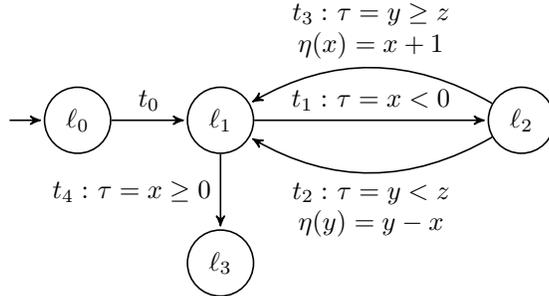

\vspace*{-.9cm}

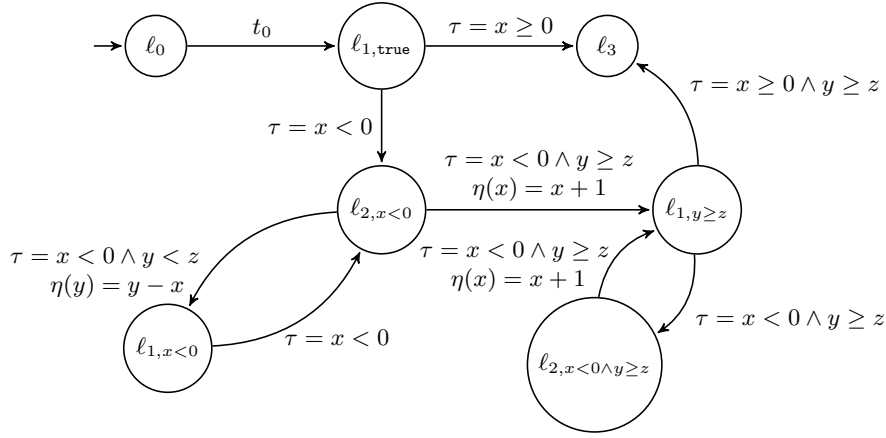
\begin{figure}[t]
	\centering
	\begin{tikzpicture}[->,>=stealth',shorten >=1pt,auto,node distance=3cm,semithick,initial text=$ $]
		\node[state,initial left] (l0) {$\location_0$};
		\node[state] (l1) [right = 2cm of l0]{$\location_{1,\true}$};
		\node[state] (l2) [below = 1cm of l1]{$\location_{2,x<0}$};
		\node[state] (l4) [right = 3cm of l2]{$\location_{1,y\geq z}$};
		\node[state] (l5) [below left = 1cm and 2cm of l2]{$\location_{1,x<0}$};
		\node[state] (l31) [right = 2cm of l1]{$\location_3$};
		\node[state] (l21) [below left = 1cm and .3cm of l4]{$\location_{2,x<0 \land y \geq z}$};

		\draw (l0) edge[above] node [align=center] {$t_0$} (l1) (l1) edge node [align=center,above] {$\tau = x\geq 0$} (l31) (l1) edge[right] node [align=center, left] {$\tau = x < 0$} (l2) (l2) edge node [align=center,above] {$\tau = x < 0\land y\geq z$\\
				$\update(x) = x + 1$} (l4) (l4) edge[bend left] node [align=center] {$\tau = x < 0\land y\geq z$} (l21) (l21) edge[bend left,below left] node [align=center,yshift=.3cm,xshift=.1cm] {$\tau = x < 0\land y\geq z\;$\\
				$\update(x) = x + 1$} (l4) (l2) edge[bend right] node [align=center,below left,xshift=-.2cm] {$\tau = x < 0 \land y < z\quad$\\
				$\update(y) = y - x$} (l5) (l5) edge[bend right,near start] node [align=center,right,xshift=.2cm] {$\tau = x < 0$} (l2) (l4) edge[bend right] node [align=center,above right]
			{$\tau = x \geq 0 \land y\geq z$} (l31);
	\end{tikzpicture}
	\caption{Applying Partial Evaluation
				to \cref{fig:cfr-running}}\label{fig:applying-pe}
                                \vspace*{-.3cm}
\end{figure}

\begin{example}
	\label{ex:unrolling}
	\cref{fig:cfr-running} represents the program from \cref{fig:crf} in our formalism for integer programs.
	Here, we used an explicit location $\location_3$ for the end of the program to illustrate how \cref{algorithm:partial_evaluation}
	handles transitions which leave the SCC.

We apply \cref{algorithm:partial_evaluation} to the program in \cref{fig:cfr-running}
and refine the SCC $\TSet_{\mathit{SCC}} = \{t_1,t_2,t_3\}$.
The entry location is $\mathcal{E}_{\TSet_{\mathit{SCC}}}= \{\location_1\}$.
To increase readability, let $\guard_i$ be the guard and $\update_i$ be the update of
transition $t_i$ for all $0 \leq i \leq 3$.

For the abstraction layers, we choose\footnote{In \cite{domenech2019control}, different heuristics are presented to choose such abstraction layers.
	In our implementation, we use these heuristics as a black box.}
$\alpha_{\location_1} = \alpha_{\location_2} = \{
	x<0, y \geq z\}$.
It is not necessary to define abstraction layers for $\location_0$ and $\location_3$, as they are not part of the SCC.
So for any constraint $\varphi_{\mathit{new}}$ and $i \in \{1,2\}$, $\alpha_{\location_i}(\varphi_{\mathit{new}})$ can only be a conjunction of the inequations in $\alpha_{\location_i}$ (i.e., $\alpha_{\location_i}(\varphi_{\mathit{new}})$ is $\true$, $x < 0$, $y \geq z$, or $x < 0 \land y \geq z$, where $\true$ corresponds to the empty conjunction).

Since $t_0$ is the only entry transition to the entry location $\location_1$, we initialize $\TSet_{\mathit{res}}$ with $\{(\location_0,\guard_0,$ $\update_0, \langle \location_1, \true\rangle)\}$ and $\LSet_1$ with $\{\langle \location_1, \true\rangle\}$.

In the first iteration $\LSet_0$ only consists of $\langle \location_1, \true\rangle$.
We have two possible transitions which we can apply in $\location_1$:\ $t_1=(\location_1,\guard_1,\update_1, \location_2)\in\TSet_{\mathit{SCC}}$ or $t_4=(\location_1,\guard_4,\update_4,\location_3)\in\TSet\setminus\TSet_{\mathit{SCC}}$.
We start with transition $t_1$.
Since the update $\update_1$ is the identity, from the guard $\guard_1 = (x<0)$ we obtain the resulting constraint $\varphi_{\mathit{new}} = (x<0)$.
We apply the abstraction layer and get $\alpha_{\location_2}(x < 0) = (x < 0)$ because $ \models (x < 0) \to (x < 0)$.
Now we add the new transition
\[
	\left(\langle \location_1, \true \rangle, \true \land \guard_1,\update_1,\langle \location_2, x < 0\rangle\right)
\]
to $\TSet_{\mathit{res}}$ and $\langle \location_2, x < 0 \rangle$ to $\LSet_1$.
For transition $t_4=(\location_1,\guard_4,\update_4,\location_3)$, we update its source location and get the resulting transition
\[
	\left(\langle \location_1,\true \rangle, \true \land \guard_4,\update_4,\location_3\right)
\]
in $\TSet_{\mathit{res}}$. We add $\langle \location_1, \true\rangle$ to $\LSet_{\mathit{done}}$.
Now $\LSet_1$ consists of $\langle \location_2, x < 0\rangle$.
There are two transitions $t_2$ and $t_3$ which can be applied in $\location_2$.
For $t_2$, from the previous constraint $x<0$ and the guard $\guard_2 = (y < z)$ we can infer that after the update $\update_2(y)=y-x$ we have $x < 0 \land y < z-x$.
As the abstraction layer $\alpha_{\location_1}$ consists of $x < 0$ and $y \geq z$, we have $\alpha_{\location_1}(x < 0 \land y < z-x) = x < 0$, since $\not\models (x < 0 \land y < z-x) \to (y \geq z)$.
Thus, we add the new transition
\[
	\left(\langle \location_2, x < 0\rangle, x < 0 \land \guard_2,\update_2,\langle \location_1, x < 0\rangle\right)
\]
to $\TSet_{\mathit{res}}$ and $\langle \location_1, x < 0\rangle$ to the set $\LSet_1$.
Similarly, for $t_3$, from $x<0$ and the guard $\guard_3 = (y \geq z)$ we infer that after $\update_3(x)=x+1$ we have $x<1 \land y \geq z$.
Here, $\alpha_{\location_1}(x < 1 \land y \geq z) = y \geq z$, since $\not\models (x<1 \land y \geq z) \to (x < 0)$.
Hence, we add
\[
	\left(\langle \location_2, x < 0\rangle, x < 0 \land \guard_3,\update_3,\langle \location_1, y \geq z\rangle\right)
\]
to $\TSet_{\mathit{res}}$ and $\langle \location_1, y \geq z\rangle$ to $\LSet_1$.
So $\LSet_1$  now consists of $\langle \location_1, x < 0\rangle$ and $\langle \location_1, y \geq z\rangle$.
For $\langle \location_1, x < 0\rangle$, in the same way as before we obtain the following
new transitions:
\begin{alignat*}{3}
	(\langle \location_1, x < 0\rangle & , x < 0 \land \guard_1 &  & ,\update_1 &  & ,\langle \location_2, x < 0\rangle) \\
	(\langle \location_1, x < 0\rangle & , x < 0 \land \guard_4 &  & ,\update_4 &  & ,\location_3)
\end{alignat*}
Note that the guard $x < 0 \land \guard_4$ of the last transition is unsatisfiable.
For that reason, we always remove transitions with unsatisfiable guard after partial evaluation was applied.
For $\langle \location_1, y \geq z \rangle$, we obtain the following new transitions:
\begin{alignat*}{3}
	(\langle \location_1, y \geq z\rangle & , y \geq z \land \guard_1 &  & , \update_1 &  & ,\langle \location_2, x < 0 \land y \geq z \rangle) \\
	(\langle \location_1, y \geq z\rangle & , y \geq z \land \guard_4 &  & ,\update_4  &  & ,\location_3)
\end{alignat*}
Thus, $\LSet_1$ now consists of the new location $\langle \location_2, x < 0 \land y \geq z \rangle$.
For this location, we finally get the following new transitions:
\begin{alignat*}{3}
	(\langle \location_2, x < 0 \land y \geq z\rangle & , x < 0 \land y \geq z \land \guard_2 &  & ,\update_2  &  & , \langle \location_1, x<0 \land y \geq z \rangle ) \\
	(\langle \location_2, x < 0 \land y \geq z\rangle & , x < 0 \land y \geq z \land \guard_3 &  & , \update_3 &  & ,\langle \location_1, y \geq z \rangle)
\end{alignat*}
Since the guard $x < 0 \land y \geq z \land \guard_2$ of the penultimate transition is again unsatisfiable, it will be removed.
For that reason, then the location $\langle \location_1, x<0 \land y \geq z \rangle$ will be unreachable and will also be removed.

\cref{fig:applying-pe}	shows the refined integer program where we wrote $\location_{i,\varphi}$ instead of $\langle \location_i, \varphi \rangle$ for readability.
Moreover, transitions with unsatisfiable guard or unreachable locations were removed.
The first SCC with the locations $\langle \location_2,x<0 \rangle$ and $\langle \location_1,x<0 \rangle$ is applied \emph{before} the second SCC with the locations $\langle \location_1,y \geq z \rangle$ and $\langle \location_2,x<0 \land y \geq z\rangle$.
So we have detected that these two SCCs occur \emph{after} each other.
Indeed, the integer program in \cref{fig:applying-pe} corresponds to the one
in \cref{fig:unrolled}.
\end{example}

\cref{algorithm:partial_evaluation} is sound because partial evaluation transforms a program $\Program$ into an \emph{equivalent} program $\Program'$.
Therefore, it does not change the runtime.

\begin{definition}[Equivalence of Programs]
	\label{def:over-approx}
	Let $\Program = (\PVSet,\LSet,\location_0,\TSet)$ and $\Program' = (\PVSet,\LSet',\location_0,\TSet')$ be integer programs over $\VSet$.
	$\Program$ and $\Program'$ are \emph{equivalent} \pagebreak
        iff the following holds for all states $\valuation_0 \in \Valuation$:
	There is an evaluation $(\location_0,\valuation_0) \to^{k}_{\TSet}
		(\location,\valuation)$ for some $\valuation \in \Valuation$, some $k \in \NN$, and some $\location\in \LSet$ iff there is an evaluation $(\location_0,\valuation_0) \to^{k}_{\TSet'} (\location',\valuation)$ for the same $\valuation \in \Valuation$ and $k \in \NN$, and some location $\location'$.
\end{definition}

\begin{theorem}[Soundness of Partial Evaluation in \cref{algorithm:partial_evaluation}]
	\label{thm:corr_partialeval}
	Let $\Program = (\PVSet,\LSet,\linebreak
        \location_0,\TSet)$ be an integer program and let $\TSet_\mathit{SCC} \subseteq \TSet$ be a non-trivial SCC of the program graph.
	Let $\Program'$ be the integer program resulting from applying \cref{algorithm:partial_evaluation} to $\Program$ and $\TSet_\mathit{SCC}$.
	Then $\Program$ and $\Program'$ are equivalent.
\end{theorem}
\makeproof{thm:corr_partialeval}{
	Let $\Program' = (\PVSet,\LSet',\location_0,\TSet')$.
	First note that for every evaluation $(\location_0,\valuation_0) \to^{k}_{\TSet'} (\location',\valuation)$ there is obviously also a corresponding evaluation $(\location_0,\valuation_0) \to^{k}_{\TSet} (\location,\valuation)$.
	To obtain the evaluation with $\TSet$ one simply has to remove the labels from the locations.
	Then the claim follows because the guards of the transitions in $\TSet'$ always imply the guards of the respective original transitions in $\TSet$ and the updates of the transitions have not been modified in the transformation from $\TSet$ to $\TSet'$.

	For the other direction, we show by induction on $k \in \NN$ that for every evaluation $(\location_0,\valuation_0) \to^{k}_{\TSet} (\location,\valuation)$ there is a corresponding evaluation $(\location_0,\valuation_0) \to^{k}_{\TSet'}
		(\location',\valuation)$ where either $\location' = \location$ or $\location' = \langle \location, \varphi \rangle$ for some constraint $\varphi$ with $\valuation(\varphi) = \true$.

	In the induction base, we have $k = 0$ and the claim is trivial.
	In the induction step $k > 0$ the evaluation has the form
	\[
		(\location_0,\valuation_0)\rightarrow_{t_1}(\location_1,\valuation_1)\rightarrow_{t_2} \cdots \rightarrow_{t_{k-1}}(\location_{k-1},\valuation_{k-1}) \rightarrow_{t_k}(\location_{k},\valuation_{k})
	\]
	with $t_1, \ldots, t_k \in \TSet$.
	By the induction hypothesis, there is a corresponding evaluation
	\[
		(\location_0,\valuation_0)\rightarrow_{t_1'}(\location_1',\valuation_1) \rightarrow_{t_2'} \cdots \rightarrow_{t_{k-1}'}(\location_{k-1}',\valuation_{k-1})
	\]
	with $t_1', \ldots, t_k' \in \TSet'$ where $\location_{k-1}' = \location_{k-1}$ or $\location_{k-1}' = \langle \location_{k-1}, \varphi \rangle$ for some constraint $\varphi$ with $\valuation_{k-1}(\varphi) = \true$.
	We distinguish two cases:
	\begin{description}
		\item[Case 1: $t_{k}\not\in\TSet_\mathit{SCC}$.] If $\location_{k-1}' = \location_{k-1}$ and $\location_k \notin \mathcal{E}_{\TSet_{\mathit{SCC}}}$, then $t_k$ has not been modified in the transformation from $\Program$ to $\Program'$.
			Thus, we have the evaluation $(\location_0,\valuation_0)\rightarrow_{t_1'}(\location_1',\valuation_1) \rightarrow_{t_2'} \cdots \rightarrow_{t_{k-1}'}(\location_{k-1}',\valuation_{k-1}) = (\location_{k-1},\valuation_{k-1}) \to_{t_k} (\location_{k},\valuation_{k})$ with $t_k \in \TSet'$.

			If $\location_{k-1}' = \location_{k-1}$ and $\location_k \in \mathcal{E}_{\TSet_{\mathit{SCC}}}$, then for $t_k = (\location_{k-1},\guard,\update,\location_{k})$, we set $\location_k' = \langle \location_k, \true \rangle$ and obtain that $t_k' = (\location_{k-1},\guard,\update,\location_{k}') \in \TSet'$.
			So we get the evaluation $(\location_0,\valuation_0)\rightarrow_{t_1'}(\location_1',\valuation_1) \rightarrow_{t_2'} \cdots \rightarrow_{t_{k-1}'}(\location_{k-1}',\valuation_{k-1}) = (\location_{k-1},\valuation_{k-1}) \to_{t_k'} (\location_{k}',\valuation_{k})$.

			Finally, we regard the case $\location_{k-1}' = \langle \location_{k-1}, \varphi \rangle$ where $\valuation_{k-1}(\varphi) = \true$.
			As $t_k = (\location_{k-1},\guard,\update,\location_{k})\in\TSet\setminus\TSet_{\mathit{SCC}}$, and $\TSet_{\mathit{SCC}}$ is an SCC, there is a $t_k' = (\langle \location_{k-1}, \varphi \rangle,\linebreak
				\varphi \land \guard, \update, \location_{k}) \in \TSet'$.
			Then $(\location_0,\valuation_0)\rightarrow_{t_1'}(\location_1',\valuation_1) \rightarrow_{t_2'} \cdots \rightarrow_{t_{k-1}'}
				(\location_{k-1}',\valuation_{k-1}) = (\langle \location_{k-1}, \varphi \rangle,\valuation_{k-1}) \rightarrow_{t_{k}'}(\location_{k},\valuation_{k})$ is an evaluation with $\TSet'$.
			The evaluation step with $t_k'$ is possible, since $\valuation_{k-1}(\varphi) = \true$ and $\valuation_{k-1}(\guard) = \true$ (due to the evaluation step $(\location_{k-1},\valuation_{k-1}) \to_{t_k} (\location_{k},\valuation_{k})$).
			Note that the step with $t_k'$ also results in the state $\valuation_{k}$, because both $t_k$ and $t_k'$ have the same update $\update$.
		\item[Case 2: $t_{k}\in\TSet_\mathit{SCC}$.] Here, $\location_{k-1}'$ has the form $\langle \location_{k-1}, \varphi \rangle$ where $\valuation_{k-1}(\varphi) = \true$.
			As $\location_k$ is part of the SCC and hence has an incoming transition from $\TSet_\mathit{SCC}$, at some point it is refined by \cref{algorithm:partial_evaluation}.
			Thus, for $t_k = (\location_{k-1}, \guard,\update, \location_{k})$, there is some $t_{k}' = \left(\langle \location_{k-1}, \varphi \rangle,\varphi \land \guard,\update, \left\langle\location_{k},\alpha_{\location_{k}}(\varphi_{\mathit{new}})\right\rangle\right) \in \TSet'$ where $\alpha_{\location_{k}}(\varphi_{\mathit{new}})$ is constructed as in \cref{line:refine}.
			This leads to the corresponding evaluation $(\location_0,\valuation_0)\rightarrow_{t_1'}(\location_1',\valuation_1) \rightarrow_{t_2'} \cdots \rightarrow_{t_{k-1}'}(\langle \location_{k-1}, \varphi \rangle,\valuation_{k-1})\rightarrow_{t_{k}'}(\langle \location_{k}, \alpha_{\location_{k}}(\varphi_{\mathit{new}})\rangle,\valuation_{k})$.
			Again, the evaluation step with $t_k'$ is possible, since $\valuation_{k-1}(\varphi) = \true$ and $\valuation_{k-1}(\guard) = \true$ (due to the evaluation step $(\location_{k-1},\valuation_{k-1}) \to_{t_k} (\location_{k},\valuation_{k})$).
			And again, the step with $t_k'$ also results in the state $\valuation_{k}$, because both $t_k$ and $t_k'$ have the same update $\update$.
			Finally, note that we have $\valuation_k(\alpha_{\location_{k}}(\varphi_{\mathit{new}})) = \true$.
			The reason is that $\models (\varphi \land \guard) \to \update(\varphi_{\mathit{new}})$ and $\valuation_{k-1}(\varphi\land \guard)= \true$ implies $\valuation_{k-1}(\update(\varphi_{\mathit{new}})) = \true$.
			Hence, we also have $\valuation_{k}(\varphi_{\mathit{new}}) = \valuation_{k-1}(\update(\varphi_{\mathit{new}})) = \true$.
			Therefore, $\models \varphi_{\mathit{new}} \to\alpha_{\location_{k}}(\varphi_{\mathit{new}})$ implies $\valuation_{k}(\alpha_{\location_{k}}(\varphi_{\mathit{new}}))= \true$.
			\qed
	\end{description}
}

\subsection{Sub-SCC-Based Partial Evaluation}
\label{sec-Local Control-Flow Refinement}

As control-flow refinement may lead to an exponential blow-up of
the program, we now present an algorithm where we heuristically minimize the strongly
connected part of the program on which we apply partial evaluation
(\cref{algorithm:cfr-local}) and we discuss how to integrate it into our approach for
complexity analysis.
Our experiments in \cref{sec:eval} show that such a sub-SCC-based partial evaluation
leads to significantly shorter runtimes than the SCC-based partial evaluation
of \cref{algorithm:partial_evaluation}.

\begin{figure}[t]
\begin{algorithm}[H]
	\SetAlgoLined
	\DontPrintSemicolon
	\KwIn{A program $\Program = (\PVSet,\LSet,\location_0,\TSet)$
		and a non-empty subset $\TSet_{\mathit{cfr}}$ of a non-trivial SCC from $\TSet$.}
	$\Set \gets \emptyset$\;
	\caption{Partial Evaluation for a Subset
		of an SCC}\label{algorithm:cfr-local}
	\ForAll{$t = ( \location,\guard, \update, \location') \in \TSet_{\mathit{cfr}}$\label{line:foreach_transition_begin}}{
		$\TSet_t {}\gets{}$ a shortest path from
		$\location'$ to $\location$\;
		$\TSet_t \gets \TSet_t \cup \{t\}$\;
		$\TSet_t {}\gets{} \TSet_t \cup \{(\hat{\location},\_,\_,\hat{\location}') \in \TSet \mid (\hat{\location},\_,\_,\hat{\location}') \in \TSet_t\}$\label{line:add_parallel}\;
		\ForAll{entry transitions
			$(\overline{\location},\overline{\guard}, \overline{\update}, \overline{\location}') \in \mathcal{ET}_{\TSet_t}$\label{line:foreach_redirect_begin}}{
			Add transition $(\location_{\mathit{new}},\overline{\guard}, \overline{\update}, \overline{\location}')$ to $\TSet_t$.\label{line:foreach_redirect_end}\;
		}
		$\Set \gets \Set \cup \{(\PVSet,\LSet,\location_{\mathit{new}},\TSet_t)\}$\label{line:foreach_transition_end}\;
	}
	\Repeat{$\Set$ does not change anymore\label{line:repeat_merge_end}}{\label{line:repeat_merge_begin}
		\If {there exist
			$\Program'=(\PVSet,\LSet,\location_{\mathit{new}},\TSet')$ and
			$\Program''=(\PVSet,\LSet,\location_{\mathit{new}},\TSet'')$
			with $\Program',\Program''\in\Set$, $\Program' \neq \Program''$, and
			a location $\location \neq \location_{\mathit{new}}$ occurs in both $\TSet'$ and $\TSet''$}{
			$\Set \gets (\Set \setminus \{\Program',\Program''\}) \cup \{(\PVSet,\LSet,\location_{\mathit{new}},\TSet'\cup \TSet'')\} $\;
		}
	}
	\ForAll{$\Program' = (\PVSet,\LSet,\location_{\mathit{new}},\TSet') \in \Set$\label{line:foreach_peri_begin}}{
		$\Program''= (\PVSet,\LSet'',\location_{\mathit{new}},\TSet'') {}\gets{}$
		apply \cref{algorithm:partial_evaluation} to $\Program'$ and the single
		non-trivial SCC $\TSet_\mathit{SCC}$ in $\TSet'$\;
		Extend the transitions $\TSet$ of $\Program$ by the transitions $\TSet''$.\;
		\ForAll{entry transitions $t
				=(\location,\guard, \update, \location')
				\in \mathcal{ET}_{\TSet'}$}{
			Replace $t$ by $(\location,\guard, \update, \langle \location', \true \rangle)$ in $\Program$.\;
		}
		\ForAll{outgoing transitions $t
				= ( \location,\guard, \update, \location') \in \mathcal{ET}_{\TSet \setminus \TSet'}$}{
			Replace $t$ by
			$( \langle \location,\varphi \rangle, \guard, \update, \location')$ in
			$\Program$
			for all $ \langle \location,\varphi \rangle\in\LSet''$.\;
		}\label{line:foreach_peri_end}
	}
	Remove unreachable locations and transitions, and transitions with
	unsatisfiable guard.\;
	\KwOut{Refined program $\Program$ \medskip}
\end{algorithm}
\vspace*{-.4cm}
\end{figure}

The idea of \cref{algorithm:cfr-local} is to find a minimal cycle of the program graph
containing the transitions $\TSet_{\mathit{cfr}}$ whose \pagebreak runtime bound we aim to improve by partial evaluation.
On the one hand,  in this way we minimize the input set $\TSet_\mathit{SCC}$ for the partial evaluation algorithm.
On the other hand, we keep enough of the original program's control flow such that partial evaluation can produce useful results.

Our local control-flow refinement technique in \cref{algorithm:cfr-local} consists of three parts.
In the first loop in \crefrange{line:foreach_transition_begin}{line:foreach_transition_end}, we find a minimal cycle $\TSet_t$ for each transition $t$ from $\TSet_{\mathit{cfr}}$.
Afterwards, $\TSet_t$ is extended by all transitions which are parallel to some transition in $\TSet_t$ in \cref{line:add_parallel}.
Otherwise, we would not be able to correctly insert the refined program afterwards.
We add a fresh initial location $\location_{\mathit{new}}$, take all entry transitions to the previously computed cycle and extend $\TSet_t$ by new corresponding entry transitions which start in $\location_{\mathit{new}}$ instead (\cref{line:foreach_redirect_begin,line:foreach_redirect_end}).
We collect all these programs in a set $\Set$, where the programs have $\location_{\mathit{new}}$ as their initial location.

So for our example from \cref{fig:cfr-running} and $\TSet_{\mathit{cfr}} = \{ t_3 \}$, $\Set$ only contains one program with locations $\location_1, \location_2, \location_{\mathit{new}}$, transitions $t_1, t_2, t_3$, and a transition from $\location_{\mathit{new}}$ to $\location_1$.

As the next step, in the second loop in \crefrange{line:repeat_merge_begin}{line:repeat_merge_end}, we merge those programs which share a location other than $\location_{\mathit{new}}$.
Again, this allows us to correctly insert the refined program afterwards (see the proof of \cref{thm:correctness_local}).

The last loop in \crefrange{line:foreach_peri_begin}{line:foreach_peri_end} performs partial evaluation on each strongly connected part of the programs in $\Set$, and inserts the refined programs into the original one by redirecting the entry and the outgoing transitions.
Here, an outgoing transition is simply an entry transition of the complement.

At the end of \cref{algorithm:cfr-local}, one should remove unreachable locations and transitions, as well as transitions with unsatisfiable guard.
This is needed, because the refined\linebreak
transitions $\TSet'$ are simply added to the old transitions $\TSet$, and entry and outgoing transitions are redirected.
So the previous transitions might become unreachable.

Instead of implementing \cref{algorithm:partial_evaluation} ourselves, our complexity
analyzer \tool{KoAT} calls the implementation of \cite{domenech2019control} in the
tool \tool{iRankFinder} \cite{irank2018} as a backend for partial evaluation.\footnote{To
ensure the \emph{equivalence}
of the transformed program according to \Cref{def:over-approx}, we call \tool{iRankFinder} with a flag to
prevent the ``chaining'' of transitions. This ensures that partial evaluation does not
change the lengths of evaluations.}
So in particular, we rely on \tool{iRankFinder}'s heuristics to compute the abstraction layers $\alpha_{\location'}$ and the new constraints $\varphi_{\mathit{new}}$ resp.\ $\alpha_{\location'}(\varphi_{\mathit{new}})$ in \cref{algorithm:partial_evaluation}.

So in our example, partial evaluation on the program in $\Set$ would result in a program like the one in \cref{fig:applying-pe}, but instead of the transition from $\location_0$ to $\location_{1,\true}$ there would be a transition from $\location_{\mathit{new}}$ to $\location_{1,\true}$.
Moreover, the location $\location_3$ and the transitions to $\location_3$ would be missing.
The redirection of the entry and the outgoing transitions would finally create the program from \cref{fig:applying-pe}.

The advantage of our technique in contrast to the na\"ive approach (i.e., applying partial evaluation on the full program as a preprocessing step) and also to the SCC-based approach in \cref{algorithm:partial_evaluation}, is that \cref{algorithm:cfr-local} allows us to apply partial evaluation ``on-demand'' just on those transitions where our bounds are still ``improvable''.
Thus, to integrate partial evaluation into our overall approach, \cref{algorithm:mprf} is modified such that after the treatment of an SCC $\widetilde{\TSet}$ in \crefrange{forall-outerLoopStart}{line:size}, we let $\TSet_{\mathit{cfr}}$ consists of all transitions $t \in \widetilde{\TSet}$ where $\Time(t)$ is not linear (and not constant).
So this is our heuristic to detect transitions with ``improvable'' bounds.
If $\TSet_{\mathit{cfr}} \neq \emptyset$, then we call \cref{algorithm:cfr-local} to
perform partial evaluation \pagebreak and afterwards we
execute \crefrange{forall-outerLoopStart}{line:size} of  \cref{algorithm:mprf}
once more for the SCC that results from refining $\widetilde{\TSet}$.

\begin{theorem}[Soundness of Partial Evaluation in \cref{algorithm:cfr-local}]
	\label{thm:correctness_local}
	Let $\Program = (\PVSet,\LSet,\linebreak \location_0,\TSet)$ be a program and $\TSet_{\mathit{cfr}}\subseteq\TSet$ a non-empty set of transitions from some non-trivial SCC.
	Then $\Program$ and the program computed by \cref{algorithm:cfr-local} are equivalent.
\end{theorem}
\makeproof{thm:correctness_local}{
	Let $\Program' = (\PVSet,\LSet',\location_0,\TSet')$ result from $\Program$ by \cref{algorithm:cfr-local}.
	As in the proof of \cref{thm:corr_partialeval}, for every evaluation $(\location_0,\valuation_0) \to^{k}_{\TSet'} (\location',\valuation)$ there is also a corresponding evaluation $(\location_0,\valuation_0) \to^{k}_{\TSet} (\location,\valuation)$, which is obtained by removing the labels from the locations.

	For the other direction, we show that for each evaluation
	$(\location_0,\valuation_0)\rightarrow_{t_1}(\location_1,\valuation_1)\rightarrow_{t_2} \cdots \rightarrow_{t_k}(\location_k,\valuation_k)$
	with $t_1,\ldots, t_k \in \TSet$ there is a corresponding evaluation
$(\location_0,\valuation_0)\rightarrow_{\TSet'}^k (\location_k',\valuation_k)$
in $\Program'$.
	To obtain this evaluation, we handle all evaluation fragments separately which use programs $\mathcal{Q}$ from $\Set$.
	This is possible, since different programs in $\Set$ do \emph{not} share locations, i.e., entry and outgoing transitions of $\mathcal{Q}$ cannot be part of another $\mathcal{Q}'$ from $\Set$.
	Such an evaluation fragment has the form
	\begin{align}
		(\location_i,\valuation_i)\rightarrow_{t_{i+1}}(\location_{i+1},\valuation_{i+1}) \rightarrow_{t_{i+2}} \cdots \rightarrow_{t_{n-1}}(\location_{n-1},\valuation_{n-1}) \rightarrow_{t_n}(\location_n,\valuation_n)\label{proof:cfr_evaluation0}
	\end{align}
	where $t_{i+1}$ is an entry transition to $\mathcal{Q}$, $t_n$ is an outgoing transition from $\mathcal{Q}$, and
	the transitions $t_{i+2}, \ldots, t_{n-1}$ belong to $\mathcal{Q}$.
	By \cref{thm:corr_partialeval} it follows that there is a corresponding evaluation using the transitions $t_{i+2}', \ldots, t_{n-1}'$ from the refined version of $\mathcal{Q}$, such that with the new redirected entry transition $t_{i+1}'$ and the new redirected outgoing transition $t_n'$ we have
	\begin{align}
		(\location_i,\valuation_i)\rightarrow_{t_{i+1}'}(\location_{i+1}',\valuation_{i+1}) \rightarrow_{t_{i+2}'}
		\cdots \rightarrow_{t_{n - 1}'}(\location_{n - 1}',\valuation_{n - 1}) \rightarrow_{t_{n}'}(\location_n,\valuation_n)\label{proof:cfr_evaluation1}
	\end{align}
	Thus, by substituting each evaluation fragment \eqref{proof:cfr_evaluation0}
	in an evaluation of $\Program$ by its refinement \eqref{proof:cfr_evaluation1}, we get a corresponding evaluation in $\Program'$.
	\qed
}

Both $\MRFs$ and control-flow refinement detect ``phases'' of the program.
An $\MRF$ represents these phases via different ranking functions, whereas control-flow refinement makes these phases explicit by modifying the program, e.g., by splitting an SCC into several new ones as in \cref{ex:unrolling}.
\cref{MPRBetterThanCFR,CFRBetterThanMPR} show that there are programs where one of the techniques allows us to infer a finite bound on the runtime complexity while the other one does not.
This is also demonstrated by our experiments with different configurations of \KoAT\ in \cref{sec:eval}.

\begin{example}
	\label{MPRBetterThanCFR}
	For the program corresponding to the loop in \cref{fig:linear} we can only infer a finite runtime bound if we search for $\MRFs$ of \emph{at least} depth $2$.
	In contrast, control-flow refinement via partial evaluation does not help here,
	because it does not change the loop.
	The used $\MRF$ $(f_1,f_2)$ with $f_1(\location_1) = f_1(\location_2) = y+1$ and
        $f_2(\location_1) = f_2(\location_2) = x$ (see \cref{ex:running_mrf}) corresponds \emph{implicitly} to the case analysis $y \geq 0$ resp.\ $y < 0$.
	However, this case analysis is not detected by \cref{algorithm:partial_evaluation}, because $y < 0$ only holds after $|y_0|+1$ executions of this loop if we have $y = y_0$ initially.
	Thus, this cannot be inferred when evaluating the loop partially for a finite number of times (as this number depends on the initial values of the variables).
	As \cref{fig:linear} does not admit a linear ranking function, this means that we fail to infer a finite runtime bound if we only use linear ranking functions and control-flow refinement.
	The same argument explains why we cannot infer a finite runtime bound for our running example in \cref{fig:running} (which contains the loop in \cref{fig:linear}) with only linear ranking functions and control-flow refinement.
	For this example, we again need $\MRFs$ of at least depth 2 (see \cref{ex:mrf_final}).
	So control-flow refinement via partial evaluation does not subsume $\MRFs$.
\end{example}
\begin{example}
	\label{CFRBetterThanMPR}
	Now we show an example where $\MRFs$ are not strong enough to infer a finite runtime bound, whereas this is possible using just linear ranking functions (i.e., $\MRFs$ of depth 1) if we apply partial evaluation before.
	Moreover, it illustrates \cref{algorithm:cfr-local}
	which only performs partial evaluation on a subset of an SCC.

	\begin{figure}[t]
		\begin{minipage}{0.4\linewidth}
			\centering \scalebox{0.8}{
				\begin{tikzpicture}[->,>=stealth',shorten >=1pt,auto,node distance=6cm,semithick,initial text=$ $]
					\node[state,initial left] (l0) {$\location_0$}; \node[state] (l1) [below = 1cm of l0]{$\location_1$}; \node[state] (l2) [below = 2cm of l1]{$\location_2$};

					\draw (l0) edge[above] node [align=center,right] {$t_0 : \update(x) = u$} (l1) (l1) edge [loop left] node [align=center,left] {$t_1:\guard = 1\leq x\leq 3$\\
							$\qquad\;\;\wedge\; w = 0$\\
							$\update(x) = x + 1$} (l1) (l1) edge [bend left, above] node [align=center,right] {$t_3:\guard = y> 0$ \\$\qquad\;\;\wedge\; w = 1$} (l2) (l2) edge [bend left, below] node [align=center,left] {$t_2:\update(y) = y - 1$} (l1);
				\end{tikzpicture}
			}
			\captionof{figure}{Original Program}\label{fig:cfr_vs_mprf_original}
		\end{minipage}
		\begin{minipage}{0.6\linewidth}
			\centering \scalebox{0.8}{
				\begin{tikzpicture}[->,>=stealth',shorten >=1pt,auto,node distance=6cm,semithick,initial text=$ $]
					\node[state,initial left] (l0) {$\location_0$}; \node[state] (l1) [below = 1cm of l0]{$\location_1$}; \node[state] (helper) [below = 1cm of l1, draw=none] {}; \node[state] (l1a) [left = 4cm of helper]{$\location_{1a}$}; \node[state] (l2) [below = 2cm of l1]{$\location_2$};

					\draw (l0) edge[above] node [align=center,right] {$t_0: \update(x) = u$} (l1) (l1) edge node [align=center,left,yshift=1cm,xshift=1.9cm]
						{$t_{1a}:\guard = 1 \leq x \leq 3$\\
						$\qquad\quad\wedge\; w = 0$\\
							$\update(x) = x+1$} (l1a) (l1) edge [bend left, above] node [align=center,right] {$t_{3a}:\guard = y > 0$\\$\qquad\quad\wedge\; w = 1$} (l2) (l1a) edge node [sloped,anchor=south,auto=false,below] {$t_3:\guard = y > 0 \wedge w = 1$} (l2) (l1a) edge [loop above] node [align=center,above] {$t_1:\guard = 2\leq x\leq 3$\\
							$\qquad\;\;\;\wedge\; w = 0$\\
							$\update(x) = x + 1$} (l1a) (l2) edge [bend left, below] node [align=center,left] {$t_2: \update(y) = y-1$} (l1);
				\end{tikzpicture}
			}
			\captionof{figure}{Result of \Cref{algorithm:cfr-local} with $\TSet_{\mathit{cfr}} = \braced{t_1}$}\label{fig:cfr_vs_mprf_transformed}
		\end{minipage}
                \vspace*{-.4cm}
	\end{figure}

	Consider the program in \cref{fig:cfr_vs_mprf_original} where $\PVSet = \{ x, y \}$ are the program variables and $\TVSet = \{u,w\}$ are the temporary variables.
	It has two independent components (the self-loop $t_1$ at location $\location_1$ and the cycle of $t_2$ and $t_3$ between $\location_1$ and $\location_2$) which do not influence each other, since $t_1$ operates only on the variable $x$ and the cycle of $t_2$ and $t_3$ depends only on $y$.
	The choice which component is evaluated is non-deterministic since it depends on
	the value of the temporary variable $w$.
	Since the value of $x$ is between $1$ and $3$ in the self-loop, $t_1$ is only evaluated at most $3$ times.
	Similarly, $t_2$ and $t_3$ are each executed at most $y$ times.
	Hence, the runtime complexity of the program is at most $1 + 3 + 2\cdot y$ = $4 + 2\cdot y$.

	However, our approach does not find a finite runtime bound when using only $\MRFs$ \emph{without} control-flow refinement.
	To make the transition $t_1$ in the self-loop decreasing, we need an $\MRF$ $f$ where the variable $x$ occurs in at least one function $f_i$ of the $\MRF$.
	So \pagebreak $f_i(\location_1)$ contains $x$ and thus, $\beta_{\location_1}$ (as defined in \cref{thm:time_bound}) contains $x$ as well.
	When constructing the global bound $\Time(t_1)$ by \cref{thm:time_bound}, we have
	to instantiate $x$ in $\beta_{\location_1}$ by $\Size(t_0,x)$, i.e., by the size-bound for $x$ of the entry transition $t_0$.
Since $x$ is set to an arbitrary integer value $u$ non-deterministically, its size is unbounded, i.e., $\Size(t_0,x) = \omega$.
	Thus, \cref{thm:time_bound} yields $\Time(t_1) = \omega$.
	The alternative solution of turning $t_0$ into a non-initial transition and adding it to the subset $\TSet'$ in \cref{thm:time_bound} does not work either.
	Since the value of $x$ after $t_0$ is an arbitrary integer, $t_0$ violates the requirement of being non-increasing for every $\MRF$ where $f_i(\location_1)$ contains $x$.

	In this example, only the self-loop $t_1$ is problematic for the computation of runtime bounds.
	We can directly infer a linear runtime bound for all other transitions, using just linear ranking functions.
	Thus, when applying control-flow refinement via partial evaluation, according to our heuristic we call \cref{algorithm:cfr-local} on just $\TSet_{\mathit{cfr}} = \braced{t_1}$.
	The result of \Cref{algorithm:cfr-local} is presented in \cref{fig:cfr_vs_mprf_transformed}.
	Since partial evaluation is restricted to the problematic transition $t_1$, the other transitions $t_2$ and $t_3$ in the SCC remain unaffected, which avoids a too large increase of the program.

	As before, in the program of \cref{fig:cfr_vs_mprf_transformed}
	we infer linear runtime bounds for $t_0, t_2, t_3$, and $t_{3a}$ using linear ranking functions.
	To obtain linear bounds for $t_{1a}$ and $t_1$, we can now use the following $\MRF$ $f$ of depth $1$ for the subset $\TSet' = \{ t_1, t_{1a}, t_3, t_{3a} \}$ and the decreasing transition $\TSet'_{>} = \{t_{1a} \}$ resp.\ $\TSet'_{>} = \{t_{1} \}$:
	\[
		f(\location_1) = 3\qquad	f(\location_{1a}) = 3-x \qquad	f(\location_2) = 0
	\]
	Thus, while this example cannot be solved by $\MRFs$, we can indeed infer linear runtime bounds when using control-flow refinement and just linear ranking functions.
	Hence, $\MRFs$ do not subsume control-flow refinement.
\end{example}

\FloatBarrier
\section{Evaluation}
\label{sec:eval}
As mentioned, we implemented both \cref{algorithm:cfr-local} and the refined version of \cref{algorithm:mprf} which calls \cref{algorithm:cfr-local}
in a new re-implementation of our tool \tool{KoAT} which is written in \tool{OCaml}.
To find $\MRFs$, it uses the SMT Solver \tool{Z3} \cite{moura2008} and it uses the tool \tool{iRankFinder} \cite{irank2018} for the implementation of \cref{algorithm:partial_evaluation} to perform partial evaluation.

To distinguish our re-implementation of \tool{KoAT} from the original version of the\linebreak
tool from \cite{koat}, let \tool{KoAT1} refer to the tool from \cite{koat} and let \tool{KoAT2} refer to our new re-implementation.
We now evaluate \tool{KoAT2} in comparison to the main other state-of-the-art tools for complexity analysis of integer programs: \tool{CoFloCo} \cite{cofloco2,cofloco3}, \tool{KoAT1} \cite{koat}, \tool{Loopus} \cite{loopus}, and \tool{MaxCore} \cite{maxcore}.
Moreover, we also evaluate the performance of \tool{KoAT1} and \tool{KoAT2} when control-flow refinement using \tool{iRankFinder} \cite{irank2018} is performed on the complete program as a preprocessing step.
We do not compare with \tool{RaML} \cite{ramlpopl17}, as it does not support programs whose complexity depends on (possibly negative) integers (see \cite{ramlweb}).
We also do not compare with \tool{PUBS} \cite{pubs}, because as stated in \cite{domenech2019control} by one of the authors of \tool{PUBS}, \tool{CoFloCo}
is stronger than \tool{PUBS}.
Note that \tool{MaxCore} is a tool chain which preprocesses the input program and then passes it to either \tool{CoFloCo} or \tool{PUBS} for the computation of the bound.
As the authors' evaluation in \cite{maxcore} shows that \tool{MaxCore} with \tool{CoFloCo} as a backend is substantially stronger than with \tool{PUBS} as a backend, we only consider the former configuration and refer to it as ``\tool{MaxCore}''.

For our evaluation, we use the two sets for complexity analysis of integer programs
from the \emph{Termination Problems Data Base (TPDB)} \cite{tpdb}
that are used in the annual \emph{Termination and Complexity Competition (TermComp)} \cite{termcomp}: \CITS{} (\CITSsh), consisting of integer transition systems, and \CCINT{} (\CCINTsh), consisting of \tool{C} programs with only integer variables.
The integers in both\linebreak
\CITSsh{} and \CCINTsh{} are interpreted as mathematical integers (i.e., without overflows).

Both \tool{Loopus} and \tool{MaxCore} only accept \tool{C} programs as in \CCINTsh{} as input.
While it is easily possible to transform the input format of \CCINTsh{} to the input format of \CITSsh{} automatically, the other direction is not so straightforward.
Hence, we compare with \tool{Loopus} and \tool{MaxCore} only on the benchmarks from the \CCINTsh{} collection.
Our tool \tool{KoAT2} is evaluated in 7 different configurations to make the effects of both control-flow refinement and $\MRFs$ explicit:
\begin{enumerate}
	\item \tool{KoAT2} denotes the configuration which uses \cref{algorithm:mprf} with maximal depth $\mdepth$ set to $1$, i.e., we only compute linear ranking functions.
	\item \tool{CFR + KoAT2} first preprocesses the complete program by performing control-flow refinement using \tool{iRankFinder}.
	      Afterwards, the refined program is analyzed with \tool{KoAT2} where $\mdepth = 1$.
	\item \tool{KoAT2 + CFRSCC} is the configuration where control-flow refinement is applied to SCCs according to \cref{algorithm:partial_evaluation}
	      and $\mdepth = 1$.
	\item \tool{KoAT2 + CFR}
	      uses \cref{algorithm:cfr-local} instead to apply control-flow refinement on sub-SCCs and has $\mdepth =1$.
	\item \tool{KoAT2 + $\MRF5$}
	      applies \cref{algorithm:mprf} with maximal depth $\mdepth = 5$, i.e., we use $\MRFs$ with up to $5$ components, but no control-flow refinement.
	\item \tool{KoAT2 + $\MRF5$ + CFRSCC} applies control-flow refinement to SCCs (\cref{algorithm:partial_evaluation}) and uses $\mdepth = 5$.
	\item \tool{KoAT2 + $\MRF5$ + CFR} uses sub-SCC control-flow refinement (\cref{algorithm:cfr-local}) and $\MRFs$ with maximal depth $\mdepth = 5$.
\end{enumerate}

Furthermore, we evaluate the tool \tool{KoAT1} in two configurations: \tool{KoAT1}
corresponds to the standalone version, whereas for \tool{CFR + KoAT1}, the complete program is first preprocessed using control-flow refinement via the tool \tool{iRankFinder} before analyzing the resulting program with \tool{KoAT1}.
The second configuration was also used in the evaluation of \tool{iRankFinder} in \cite{domenech2019control}.

We compare the runtime bounds computed by the tools asymptotically as functions which depend on the largest initial absolute value $n$ of all program variables.
All tools were run inside an Ubuntu Docker container on a machine with an AMD Ryzen 7 3700X octa-core CPU and $32 \, \mathrm{GB}$ of RAM.
The benchmarks were evaluated in parallel such that at most 8 containers were running at once, each limited to 1.9 CPU cores.
In particular, the runtimes of the tools include the times to start and remove the container.
As in \emph{TermComp}, we applied a timeout of 5 minutes for every program.
See \cite{koatwebsite} for a binary and the source code of our tool \tool{KoAT2}, a Docker image, web interfaces to test our implementation, and full details on all our experiments in the evaluation.

\begin{figure}[t]
	\label{eval_CITS_koat}
	\begin{center}
		\makebox[\textwidth][c]{
			\begin{tabular}{l|c|c|c|c|c|c||c|c}
                                                                       & $\landau(1)$ & $\landau(n)$ & $\landau(n^2)$ & $\landau(n^{>2})$ & $\landau(\mathit{EXP})$ & $< \infty$ & $\mathrm{AVG^+(s)}$ & $\mathrm{AVG(s)}$ \\
				\hline \tool{KoAT2 + $\MRF5$ + CFR}    & 131          & 255          & 101            & 13                & 6                       & 506        & 4.31    &  26.50 \\
				\hline \mbox{\footnotesize {\tool{KoAT2$\!\,$+$\,\!\MRF5\!\,$+$\,\!$CFRSCC}}} & 131          & 255          & 102            & 12                & 6                       & 506        & 6.00    &  27.47 \\
				\hline \tool{KoAT2 + CFR}              & 131          & 245          & 101            & 10                & 6                       & 493        & 5.00    &  21.81 \\
				\hline \tool{KoAT2 + CFRSCC}           & 131          & 245          & 101            & 9                 & 6                       & 492        & 6.37    &  23.45 \\
				\hline \tool{KoAT2 + $\MRF5$}          & 126          & 235          & 100            & 13                & 6                       & 480        & 2.19    &  13.32 \\
				\hline \tool{KoAT1}                    & 132          & 214          & 104            & 14                & 5                       & 469        & 0.65    &  9.38 \\
				\hline \tool{CFR + KoAT1}              & 128          & 231          & 93             & 10                & 5                       & 467        & 5.54    &  40.81 \\
				\hline \tool{CFR + KoAT2}              & 130          & 231          & 93             & 6                 & 6                       & 466        & 10.44   &  43.05 \\
				\hline \tool{CoFloCo}                  & 126          & 231          & 95             & 9                 & 0                       & 461        & 3.44    &  18.40 \\
				\hline \tool{KoAT2}                    & 126          & 218          & 97             & 10                & 6                       & 457        & 2.29    &  9.45 \\
			\end{tabular}
		}
	\end{center}\vspace*{-.4cm}
	\caption{Evaluation on \CITS{}}
	\label{fig:eval_cits}
        \vspace*{-.4cm}
\end{figure}

\subsection{Evaluation on \CITS{}}
The set \CITSsh{} consists of 781 integer programs, where at most 564 of
them \emph{might} have finite runtime (since the
tool \tool{LoAT} \cite{lowerbounds20,loat}  proves unbounded runtime complexity for 217 examples).
 The results of our experiments on this set can be found
 in \cref{fig:eval_cits}.
So for example, there are $131 + 255 = 386$ programs where \tool{KoAT2 + $\MRF5$ + CFR}
can show that $\rc(\valuation_0) \in \landau(n)$ holds for all initial states $\valuation_0$ where $\abs{\valuation_0(v)} \leq n$ for all $v \in \PVSet$.
For $131$ of these programs, \tool{KoAT2 + $\MRF5$ + CFR} can even show that $\rc(\valuation_0) \in \landau(1)$, i.e., their runtime complexity is constant.
In \cref{fig:eval_cits}, ``$< \infty$'' is the number of examples where a finite bound on the runtime complexity could be computed by the respective tool within the time limit.
``$\mathrm{AVG^+(s)}$'' is the average runtime of the tool on successful runs in seconds, i.e., where the tool proved a finite time bound before reaching the timeout, whereas ``$\mathrm{AVG(s)}$'' is the average runtime of the tool on all runs including timeouts.

\tool{KoAT2} without $\MRFs$ and control-flow refinement infers a finite bound for 457 of the 781 examples, while \tool{CoFloCo} solves 461 and \tool{KoAT1} solves 469 examples.
In contrast to \tool{KoAT2}, both \tool{CoFloCo} and \tool{KoAT1} always apply some form of control-flow refinement.
However, \tool{KoAT1}'s control-flow refinement is weaker than the one in \Cref{sec:cfr}, since it only performs loop unrolling via  ``chaining'' to combine subsequent transitions.
Indeed, when adding control-flow refinement as a preprocessing technique (in \tool{CFR + KoAT1} and \tool{CFR + KoAT2}), the tools are almost equally powerful.

However,
for efficiency it is much better to
integrate control-flow refinement into \tool{KoAT2} as
in \cref{algorithm:partial_evaluation} or \cref{algorithm:cfr-local}
(\tool{KoAT2 + CFRSCC} resp.\ \tool{KoAT2 + CFR})
than to use it
as a preprocessing step (\tool{CFR + KoAT2}).
This integration reduces the number of timeouts and therefore increases power.
The corresponding configurations already make \tool{KoAT2} stronger than all previous tools on this benchmark.
Nevertheless, while control-flow improves power substantially, it increases the
resulting runtimes.
The reason is that partial evaluation can lead to an exponential blow-up of the program.
Moreover, we have to analyze parts of the program twice:\ we first analyze parts where we do not find a linear or a constant bound.
Then, we apply control-flow refinement and afterwards, we analyze them again.

If instead of using control-flow refinement, the maximum depth of $\MRFs$ is increased from 1 to 5, \tool{KoAT2} can compute a finite runtime bound for 480 examples.
As explained in \cref{sec:mrf}, $\MRFs$ are a proper extension of classical linear ranking functions as used in \tool{KoAT1}, for example.
Thus, \tool{CoFloCo}, \tool{KoAT1}, and \tool{KoAT2 + CFR}
fail to compute a finite bound on the runtime complexity of our running example in \cref{fig:running}, while \tool{KoAT2 + $\MRF5$}
succeeds on this example.
In particular, this shows that \tool{KoAT2 + CFR} does \emph{not} subsume \tool{KoAT2 +
$\MRF5$} but the two techniques presented in \Cref{sec:mrf,sec:cfr}
can have orthogonal effects and combining them leads to an even more powerful tool.
Indeed, \tool{KoAT2 + $\MRF5$ + CFR} proves a finite bound for more examples than \tool{KoAT2 + $\MRF5$} and \tool{KoAT2 + CFR}, in total 506.
The configuration \tool{KoAT2 + $\MRF5$ + CFRSCC} has approximately the same power, but a slightly higher runtime.

\begin{figure}[t]
	\begin{minipage}{0.45\linewidth}
		\hspace*{1cm}	\LinesNotNumbered \DontPrintSemicolon
		\begin{algorithm}[H]
			\While{$x > 0$}{
				$x \gets x + y$\;
				$y \gets y + z$\;
				$z \gets z - 1$\;
			}
		\end{algorithm}
                \vspace*{-.2cm}
		\captionof{figure}{Loop With Three Phases}\label{fig:mrf_generala}
	\end{minipage}
	\begin{minipage}{0.45\linewidth}
		\begin{tikzpicture}[->,>=stealth',shorten >=1pt,auto,node distance=2cm,semithick,initial text=$ $]
			\node[state,initial] (q1) {$\location_0$}; \node[state] (q2) [right of=q1]{$\location_1$};

			\draw (q1) edge[above] node [text width=3.5cm,align=center] {$t_0$} (q2) (q2) edge[loop right] node [text width=3cm,align=center] {$t_1:\guard = x > 0$ \\
					$\update(x) = x + y $ \\
					$\update(y) = y + z$ \\
					$\update(z) = z - 1$} (q2);

		\end{tikzpicture}\vspace*{-.2cm}
		\captionof{figure}{Integer Program}\label{fig:mrf_general}
	\end{minipage}
        \vspace*{-.4cm}
\end{figure}

\subsection{Evaluation on \CCINT{}}

The benchmark suite \CCINTsh{} consists of 484 \tool{C} programs, where 366 of them \emph{might} have finite runtime (since \tool{iRankFinder}
can show non-termination of 118 examples).
To apply \tool{KoAT1} and \tool{KoAT2} on these benchmarks, one has to translate the \tool{C} programs into integer programs as in \cref{def:integer_program}.
To this end, we use the tool \tool{llvm2kittel} \cite{loop-unrolling} which performs this translation by using an intermediate representation of \tool{C} programs as \tool{LLVM}
bytecode \cite{dblp:conf/cgo/lattnera04}, obtained from the \tool{Clang} compiler frontend \cite{clang}.
The output of this transformation is then analyzed by \tool{KoAT1}
and \tool{KoAT2}.

The results of our evaluation on \CCINTsh{} can be found in \cref{fig:eval_ccint}.
Here, \tool{Loopus} solved 239 benchmarks, \tool{KoAT2} solved 281, \tool{KoAT1} solved 285, and \tool{CoFloCo} solved 288 out of the 484 examples.
Additionally, both \tool{MaxCore} and \tool{KoAT2 + $\MRF5$} solve 310 examples and \tool{KoAT2 + CFRSCC} solves 320 examples.
This makes \tool{KoAT2} the strongest tool on both benchmark sets.
Applying partial evaluation on sub-SCCs instead of SCCs improves the average runtime of successful runs, without reducing the number of solved examples.
When enabling both control-flow refinement and multiphase-linear ranking functions
then \tool{KoAT2} is even stronger, as \tool{KoAT2 + $\MRF5$ + CFR}
solves 328 examples.
Moreover, it is faster than the equally powerful configuration \tool{KoAT2 + $\MRF5$ + CFRSCC}.

In contrast to \tool{KoAT1} and \tool{CoFloCo}, \tool{MaxCore} also proves a linear runtime bound for our example in \cref{fig:linear}, as it detects that $y$ is eventually negative.
However, when generalizing \cref{fig:linear} to three phases as in \cite{genaim_mrf}
(see \cref{fig:mrf_generala,fig:mrf_general}), \tool{KoAT2} with $\MRFs$ can infer the
finite bound $27 \cdot x + 27 \cdot y + 27 \cdot z + 56$ on the runtime by using the
$\MRF$ $(z+1, y+1, x)$, whereas the other tools fail.
Moreover, \tool{KoAT2} with $\MRFs$ is the only tool that proves a finite time bound for the program in \cref{fig:running}.
To evaluate \tool{Loopus} and \tool{MaxCore} on this example, we translated it into \tool{C}.
While these tools failed, \tool{KoAT2} also succeeded on the integer program that was obtained by applying \tool{llvm2kittel} to the translated program.
This shows the robustness of $\MRFs$ for programs consisting of several phases.

For the example in \Cref{fig:cfr_vs_mprf_original} which demonstrates that $\MRFs$ do not
subsume control-flow refinement (\cref{CFRBetterThanMPR}),
\tool{KoAT2}  with its control-flow refinement technique of \cref{sec:cfr}
infers a linear runtime bound
whereas \tool{KoAT1} fails, since its loop
un\-rolling technique is a substantially weaker
form of
control-flow refinement.
Besides our tool, only \tool{MaxCore} was able to infer a finite runtime bound for
the \tool{C} version of this program, where however this bound was quadratic instead of linear.

To sum up, both multiphase-linear ranking functions and control-flow refinement lead to significant improvements.
Combining the two techniques, our tool \tool{KoAT2} outperforms \emph{all} existing state-of-the-art tools on both benchmark sets.

\begin{figure}[t]
	\label{eval_c_complexityITS_aprove}
	\begin{center}
		\begin{tabular}{l|c|c|c|c|c|c||c|c}
                                                               & $\landau(1)$ & $\landau(n)$ & $\landau(n^2)$ & $\landau(n^{>2})$ & $\landau(\mathit{EXP})$ & $< \infty$ & $\mathrm{AVG^+(s)}$ & $\mathrm{AVG(s)}$  \\
			\hline \tool{KoAT2 + $\MRF5$ + CFR}    & 24           & 228          & 65             & 11                & 0                       & 328        & 4.77     & 16.40    \\
			\hline  \mbox{\footnotesize {\tool{KoAT2$\!\,$+$\,\!\MRF5\!\,$+$\,\!$CFRSCC}}} & 24           & 228          & 65             & 11                & 0                       & 328        & 5.72     & 16.53    \\
			\hline \tool{KoAT2 + CFR}              & 25           & 216          & 68             & 11                & 0                       & 320        & 5.14     & 11.67    \\
			\hline \tool{KoAT2 + CFRSCC}           & 28           & 216          & 66             & 10                & 0                       & 320        & 6.00     & 11.93    \\
			\hline \tool{MaxCore}                  & 23           & 214          & 66             & 7                 & 0                       & 310        & 1.94     & 5.24     \\
			\hline \tool{KoAT2 + $\MRF5$}          & 23           & 204          & 71             & 12                & 0                       & 310        & 2.11     & 5.16     \\
			\hline \tool{CFR + KoAT2}              & 27           & 200          & 70             & 2                 & 1                       & 300        & 11.26    & 19.92    \\
			\hline \tool{CFR + KoAT1}              & 29           & 187          & 74             & 7                 & 0                       & 297        & 5.34     & 12.64    \\
			\hline \tool{CoFloCo}                  & 22           & 195          & 66             & 5                 & 0                       & 288        & 0.81     & 2.95     \\
			\hline \tool{KoAT1}                    & 25           & 168          & 74             & 12                & 6                       & 285        & 2.36     & 2.97     \\
			\hline \tool{KoAT2}                    & 23           & 176          & 70             & 12                & 0                       & 281        & 2.05     & 2.76     \\
			\hline \tool{Loopus}                   & 17           & 169          & 49             & 4                 & 0                       & 239        & 0.84     & 0.72
		\end{tabular}
	\end{center}
        \vspace*{-.3cm}
	\caption{Evaluation on \CCINT{}}
	\label{fig:eval_ccint}
         \vspace*{-.4cm}
\end{figure}

\FloatBarrier
\section{Related Work and Conclusion}
\label{sec:conclusion}
\paragraph{Related Work.}
As mentioned in \Cref{sec:intro}, many other techniques for automated complexity analysis of integer programs have been developed.
The approach in \cite{rank}\linebreak
uses lexicographic combinations of linear ranking functions and Ehrhart polynomials to over-approximate the runtime complexity of integer programs.
In \cite{loopus}, difference logic is used to analyze \tool{C}
programs.
The works in \cite{pubs,pubs-upper-lower,cofloco2,cofloco3,costa-complexity}
over-approximate so-called cost relations which are closely related to recurrence
relations. 
In \cite{maxcore}, a tool chain is presented which uses conditional termination proofs as in \cite{conditional-term} to guide the inference of complexity bounds via cost relations by a complexity analyzer in the backend.
Based on tools for complexity analysis of integer programs, there also exist approaches to analyze complexity for full programming languages like \tool{Java} \cite{aprove-java,dblp:journals/iandc/mosers18}.
In this way, they complement successful tools for functional verification of \tool{Java}
programs like \cite{key}.
Other approaches use the potential method from amortized analysis or type systems to analyze the complexity of \tool{C} (see, e.g., \cite{c4b}) or \tool{ML} programs \cite{ramlpopl17,dblp:journals/toplas/0002ah12}.
An approach to verify whether a given resource bound for a program is valid is presented in \cite{campy}.
While all of these works focus on over-approximating the worst-case runtime complexity of programs,
there is also work on the inference of lower bounds on the worst-case runtime complexity, see, e.g., \cite{lowerbounds20,dblp:journals/pacmpl/wangh19,dblp:conf/cav/albertgmmr20}.
Moreover, our tool \KoAT\ also offers the possibility to analyze the expected runtime complexity of probabilistic integer programs, because we also transferred the approach from \cite{koat} to probabilistic integer programs \cite{koatprob} and we also integrated decision procedures for the termination and complexity of restricted classes of probabilistic programs in \KoAT\ \cite{dblp:conf/cade/gieslgh19}.
See \cite{foundationsexpectedruntime2020} for an overview on runtime analysis for probabilistic programs.

A fundamentally different concept to integer programs are so-called term rewrite systems.
These systems model recursion and algebraic data structures, but they do not have any built-in data type for integers.
There is also a wealth of techniques and tools to analyze the runtime complexity of term rewrite systems automatically (see \cite{dp-framework,aprove,trs-lower,dblp:conf/rta/avanzinim13,dblp:conf/tacas/avanzinims16}, for example).

Multiphase-linear ranking functions are studied in \cite{genaim_mrf,leike_linear,genaim_mrf_recurrent,dblp:journals/sttt/yuanls21},
but these works mainly focus on termination instead of complexity analysis.
Moreover, \cite{genaim_mrf} shows how to obtain a linear bound on the runtime complexity of a program with a \emph{single} $\MRF$, while we developed a technique to combine $\MRFs$ on program parts to obtain bounds on the runtime complexity of the full program.

Using control-flow refinement for inferring runtime bounds is studied in \cite{cofloco3,domenech2019control}.
Here, \cite{cofloco3} focuses on cost relations, while we embed the approach of \cite{domenech2019control} into our analysis of integer programs, where we do not apply this method globally but only locally on parts where we do not yet have a linear runtime bound.

\paragraph{Conclusion.}
In this paper, we showed how to adapt the approach for the computation of runtime and size bounds for integer programs from \cite{koat} to multiphase-linear ranking functions and to the use of control-flow refinement.
As shown by our experimental evaluation, due to these new improvements, the resulting implementation in our new version of the tool \KoAT{}
outperforms the other existing tools for complexity analysis of integer programs.

\KoAT's source code, a binary, and a Docker image
are available at
\[ \mbox{\url{https://aprove-developers.github.io/ComplexityMprfCfr/}.}\]
This web site also provides details on our experiments and \emph{web interfaces} to run \KoAT\ directly online.

\paragraph{Acknowledgments.}
This paper is dedicated to Reiner Hähnle whose ground-brea\-king results on functional
verification and symbolic execution of \textsf{Java} programs with the \tool{KeY}
tool \cite{key}, on automatic resource analysis \cite{cofloco3}, and on its combination
with deductive verification (e.g., \cite{dblp:journals/sosym/albertbghpr16}) were a
major inspiration for us.
Reiner's work motivated us to develop and improve \tool{KoAT} such that it can be used as a backend for complexity analysis of languages like \tool{Java} \cite{aprove-java}.

We are indebted to Samir Genaim and Jes\'us J.\ Dom\'enech for their help and advice with integrating multiphase-linear ranking functions and partial evaluation into our approach, and for providing us with a suitable version of \tool{iRankFinder} which we could use in \KoAT's backend.
Moreover, we are grateful to Albert Rubio and Enrique Mart\'in-Mart\'in for providing us with a static binary of \tool{MaxCore}, to Antonio Flores-Montoya and Florian Zuleger for their help in running \tool{CoFloCo} and \tool{Loopus} for our experiments, and to Florian Frohn for help and advice.

\bibliographystyle{splncs04}

\end{document}